\documentclass[structabstract]{aa}  
\usepackage{graphicx}
\usepackage{txfonts}
\usepackage{natbib}
\bibpunct{(}{)}{;}{a}{}{,} 

\usepackage{longtable}
\begin{document}
   \title{The bright galaxy population of five medium redshift clusters}
   \subtitle{I. Color-Magnitude Relation, Blue Fractions and Visual Morphology}

   \author{B. Ascaso  \inst{1}, M. Moles \inst{1},
          J. A. L. Aguerri \inst{2}, R. S\'anchez-Janssen 
          \inst{2} and J. Varela\inst{1}
          }

   \institute{Instituto de Astrof\'isica de Andaluc\'ia-CSIC, Camino
Bajo de Huetor 50, C.P: 18008 Granada, Spain\\
              \email{ascaso@iaa.es, moles@iaa.es, jesusvl@iaa.es}
         \and
             Instituto de Astrof\'isica de Canarias, C/ Via
	  L\'actea S/N C.P: La Laguna, Spain\\
             \email{jalfonso@iac.es, ruben@iac.es}
             }

   \date{Received 18 February 2008 / Accepted 28 May 2008}


  \abstract
   {}
   {Using data of five clusters of galaxies within the redshift range 0.15 $\leq$ z $\leq$ 0.25, imaged with the Nordic Optical Telescope (NOT) in the central  $\approx$ 1 Mpc$^2$ in very good seeing conditions,  we have performed an exhaustive inspection of their bright galaxy population. That range of redshift, where only a small amount of data with the required resolution and quality is available, is particularly important for the understanding of the formation and evolution of clusters of galaxies. }
   {We have inspected the color-magnitude relation (CMR) for those clusters and measured the blue fraction of galaxies in their cores to check for evidence of evolution as found in other works. Moreover, the visual classification of the galaxy morphology has been performed and the morphology-radius relation has been examined}
   {We have not found signs of evolution neither in the slope of the CMR nor in the blue fraction of galaxies. A diversity of situations regarding those parameters and in the morphological mixing has been noticed, with two out of five clusters containing a dominant late-type core population. The cluster A1878 stands out as some of its properties differ from those of the other clusters in the sample. }
   { No clear signs of evolution appear in our analysis. The data support the view that the morphology and the stellar content  of the galaxies in our clusters have been already settled at z $\sim$ 0.2. Only the fraction of interacting galaxies in the clusters appear to be larger than in clusters like Coma although the number of clusters in the sample is small to give a definitive conclusion.}


 \titlerunning{Properties of galaxies in clusters at medium redshift. I}
 \authorrunning{Ascaso et al.}
   \maketitle
%

\section{Introduction}

Clusters of galaxies are the largest structures in the Universe that are gravitationally bounded. They are crowded with hundreds to thousands of galaxies. Numerous studies up to date have been devoted to the formation of clusters of galaxies. However, two main scenarios for its clarification still remain. On one hand, we have the monolithic scenario in which the clusters were formed first (\cite{bower98}) and on the other, we have the hierarchical scenario (\cite{deLucia07b,kauffmann94}), in which the galaxies were formed at the outset. The monolithic scenario implies that the galaxies are not suffering  substantial transformations after the cluster collapse \citep{merritt84} while the hierarchical scenario would imply that the environmental effects and interactions are transforming the galaxy population due to mechanisms that were operational until recent epochs, such as harassment (\cite{moore96}), gas-stripping \citep{gunn72,quilis00}, starvation (\cite{bekki02}), or merging \citep{gerhard83,aguerri01,eliche-moral06}. 

Likewise, the evolution of the galaxy population in clusters of galaxies has been broadly  studied in many works. Some of the results found up to date are the  lower fraction of lenticular galaxies and the larger fraction of late type galaxies found in the cores of higher redshift clusters (\cite{fasano00,dressler97}). The possible evolution of the slope of the color-magnitude relation (CMR) has also been widely explored (\cite{lopezcruz04,driver06,mei06,deLucia07a}).  The agreed result is that the CMR does not change with redshift up to redshift z $\sim$ 1 at least. This feature is very interesting in itself because it gives information about the metallicity and age of the galaxy population (\cite{kodama99}).

Another attribute that is considered in the context of the evolution of clusters of galaxies is the blue fraction of the galaxy population in clusters. In the early work by \cite{butcher84}, an increase of this  blue fraction with redshift was found for clusters up to redshift $\approx$ 0.5. That fact indicates that the galaxy population would be evolving. However, as shown by \cite{aguerri07}, that variation would happen only for some redshift range. They studied a large sample of SDSS clusters up to redshift z $\leq$ 0.1 and did not see any significant change of the blue fraction with the redshift. Therefore, exploring the next redshift range,  $0.1 \lesssim z \lesssim 0.3$, would be relevant to clarify the situation. In particular, since several works have explored and noticed the Butcher-Oemler effect with samples of clusters at lower (\cite{depropris04}) and higher (\cite{deLucia07a}) redshift.

Few works have been dedicated to study the morphology of the galaxy population at z$\approx$ 0.2. The morphological studies have been generally confined to rather local samples, in part due to the need to establish a visual  classification (\cite {dressler80,fasano00}) and more generally, to the difficulties to get deep and high-resolution images for relatively large fields. Some of these studies have tried to establish an automatic morphological classification by inspecting the galaxies surface brightness and their main structural parameters. Nevertheless, those samples have often been preselected to be only late type (\cite{deJong96,graham01}) or early type (\cite{graham03}). As a consequence, the present number of clusters that have been studied in that redshift range is small (\cite{fasano00,trujillo01,fasano02}). 

That range of redshift, however, should be inspected to link the results found for local clusters and for more distant objects that can be explored with the Hubble Space Telescope (HST). There are many evidences which indicates that some properties of the clusters and their galaxies could change between the local and  $\sim$ 0.4 redshift ranges.

In the present work we have studied five clusters of galaxies in a range of redshift from 0.15 to 0.25, observed with NOT at La Palma in two bands: Gunn-r and Bessel B, under very good seeing conditions. The clusters presented in this paper were already considered by \cite{fasano00,fasano02} for their study of the morphological mixing. They found that, for galaxies with $M_{V}\leq -20$, the fraction of S0 galaxies was lower for higher z clusters. We have analyzed here the main characteristics of the bright galaxy population in the central part of those five clusters. We are interested in the possible evolution with z of properties of the galactic populations such as the CMR, the fraction of blue galaxies or the radial distribution of the different (visually determined) morphological types. We have used as templates at lower redshift the results obtained for galaxy clusters at $z \leq 0.12$ by \cite{depropris04} for comparison. The analysis of the surface brightness distribution and quantitative morphology of the galaxies will be presented in a forthcoming paper, (Ascaso et al., 2008, in preparation).

The structure of this paper is as follows. In section 2, we present the data, including a brief description of the clusters environment and explain the observations and the reduction process. In section 3, we describe how the galaxies were detected and extracted. The section 4 is devoted to the study of the color-magnitude diagrams and the Butcher-Oemler effect.  Section 5 is dedicated to the study of the morphology of the galaxies from a qualitative point of view. We have considered here several aspects such as the morphology-radius relation, the concentration parameter or the frequency of interaction systems in our sample. Finally, we present the discussion and conclusions in section 6. Throughout this paper we have adopted the standard $\Lambda$CDM cosmology with H$_0$=71 km s$^{-1}$ Mpc$^{-1}$, $\Omega_m$=0.27 and $\Omega_{\Lambda}$=0.73.


\section{Observations, data reduction and description of the selected clusters.}

Two of us (M.M. and J.A.L.A.) imaged five galaxy clusters with the 2.5m Nordic Optical Telescope (NOT) located at the Roque de Los Muchachos Observatory (La Palma). The observations were taken with the Stand Camera, with a field of view of $3^{'} \times 3^{'}$, a plate scale of 0.176$^{''}$/pix, a gain of 1.69 e$^{-}$/ADU and a readout noise of 6.36 e$^{-}$. All images were taken under photometric sky conditions and very good seeing (between 0.5 $^{''}$ and 0.8$^{''}$). 

The clusters were observed through two broad-band filters: Gunn-r$^{'}$ (\emph{r}) and Bessel B (\emph{B}). At least two exposures for each field in every filter were usually taken, allowing us to clean-up the combined images for cosmic-rays and spurious events. The data reduction and calibration was  performed as presented in \cite{fasano02}. Here we summarize the basic steps of the data reduction process, for more information we refer the reader to \cite{fasano02}. 

The bulk of the data reduction of the images was achieved using standard IRAF tasks. The electronic bias level was removed from the CCD by fitting a Chebyshev function to the overscan region and subtracting it from each column. By averaging ten bias frames, a master bias per night was created and subtracted from the images in order to remove any remaining bias structure. Dark images were also observed in order to remove the dark signal from the CCD. This correction turned out to be negligible, and was not considered. Twilight flats were also observed at the beginning and at the end of every observing night. They were combined and used for removing the pixel-to-pixel structure of the images. 

The photometric calibration of the images was obtained by observing several standard stars from the \cite{landolt}, \cite{jorgensen94}, and \cite{montgomery} catalogues. They were observed every night at different zenith distances in order to measure the atmospheric extinction. The calibration constant was taken from \cite{fasano02}.
 
     \subsection{Clusters Environment}

Given the scarce information existing on the clusters presented here, we comment briefly  the redshift data and the environmental situation of each of them. 

{\bf A1643}. The redshift of this cluster is attributed by \cite{humason56}, who obtained a spectrum of the brightest galaxy in the area, finding z = 0.198. Our images were centered at that position, $\alpha(J2000)$=12h55m54.4s, $\delta(J2000)$= +44d04m46s. More recently, \cite{gal03} detected an overdense region centered at $\alpha(J2000)$=12h55m42.4s, $\delta(J2000)$=  +44d05m22s, identified as a cluster designed by NSC J125542+440522. They have determined a photometric redshift of 0.2515. Both clusters do appear in our frames where we can identify A1643 as the one dominated by the galaxy observed by \cite{humason56} and, therefore, at z = 0.198. This is the value we adopt in this paper. We will exclude the frames that could be contaminated by the presence of NSC J125542+440522 in all the analysis regarding the galactic content of A1643.

{\bf A1878}. This clusters appears with z = 0.254 in the NED \footnote{NASA/IPAC  Extragalactic Database}. A closer inspection shows that there is another value given to a galaxy in the field, namely z = 0.222. Both redshift values come from \cite{sandage76}, who observed the brightest galaxy in the field, placed at $\alpha(J2000)$=14h12m52.13s, $\delta(J2000)$=  +29d14m29s, and another, fainter galaxy at $\alpha(J2000)$=14h12m49.13s, $\delta(J2000)$=  +29d12m59s. As quoted by the authors, the spectra were of low quality. The low z value corresponds to the brightest object that appears at the center of a strong concentration of galaxies that do correspond to the cluster catalogued as A1878. More recently, \cite{gal03}, identified a cluster labeled as NSCJ141257+291256, with a photometric redshift z = 0.22. Its position and redshift value coincide with that of the bright galaxy observed by \cite{sandage76} that is accepted here as the brightest galaxy of A1878.

{\bf A1952}. The redshift attributed to this cluster, z = 0.248, also comes from the work by \cite{sandage76}  who observed the brightest cluster galaxy. The possible confusion regarding this cluster comes from the fact that the position given by \cite{abell89}, $\alpha(J2000)$=14h41m04.2s, $\delta(J2000)$=  +28d38m12s, does not coincide with that of its  Brightest Cluster Galaxy (BCG) as given by \cite{sandage76}, $\alpha(J2000)$=14h41m03.6s, $\delta(J2000)$=  +28d36m59.68s.  To add to the confusion, \cite{gal03} detected a cluster designed by NSC J144103+283622, at almost exactly the position of A1952's BCG, but the redshift they have determined photometrically amounts to 0.2084.  Taking all the information at hand, we consider that the cluster identified by \cite{gal03} is A1952, but the redshift we adopt here is that measured by \cite{sandage76}, z = 0.248.  The analysis we present here of the Color-Magnitude Relation support this conclusion.

{\bf A2111}. That cluster has the largest amount of information available in the literature of all the clusters in our sample. The redshift was established from spectroscopic observations by \cite{lavery86}. The center given by NED comes from the ACO catalogue given by \cite{abell89}, namely, $\alpha(J2000)$=15h39m38.3s, $\delta(J2000)$=  +34d24m21s. However, the subsequent analysis of the X-ray data by \cite{wang97,henriksen99,miller06}, let them conclude that the cluster center position is at $\alpha(J2000)$=15h39m40.9s , $\delta(J2000)$= +34d25m04s, only 5.04 kpc away from the Brightest Cluster Galaxy. \cite{miller06} also provides a large number of spectra. Interestingly, the X-ray works showed that this cluster is undergoing a merger and presents a rather large blue fraction.

{\bf A2658}. The redshift of that cluster is set from  \cite{fetisova82}.  The center, as given by \cite{abell89} is at $\alpha(J2000)$=23h44m58.8s, $\delta(J2000)$=  -12d18m20s. However, our BCG is located at $\alpha(J2000)$=23h44m49.83s, $\delta(J2000)$=  -12d17m38.93.  After a visual inspection of the cluster image in the Digital Sky Survey, we conclude that the center of the cluster is given by the BCG, where a high concentration of galaxies is visually detected. 

The adopted central position and redshift value for the five clusters discussed here are collected in Table~\ref{tab:all}. We also show in the table the details of the observations. Columns 1, 2, 3 and 4 give the cluster name, center coordinates (right ascension and declination in Equatorial coordinates, J2000) and redshift, respectively. The number of pointings observed for each cluster are indicated in column 5. These pointings cover different cluster areas showed in column 6. The last column gives the seeing of the observed images. For the center position we have adopted that of their respective BCGs except for A2111, for which the X-ray data has been used to set it. 

 \begin{table*}[h]
      \caption[]{The sample of Clusters}
      \[
         \begin{array}{llllllllccclll}
            \hline\noalign{\smallskip}
\multicolumn{1}{c}{\rm Name}&
\multicolumn{3}{c}{\alpha (2000)}&
\multicolumn{3}{c}{\delta (2000)}&
\multicolumn{1}{c}{z}&
\multicolumn{1}{c}{\# frames}&
\multicolumn{1}{c}{\rm Area (Mpc}$^2$)&
\multicolumn{1}{c}{\rm seeing \, ('')}\\
\hline\noalign{\smallskip}
{\rm A~2658} & 23 & 44 & 49 & -12 & 17 & 39 & 0.185 & 1 & 0.3055   & 0.70 \\
{\rm A~1643} & 12 & 55 & 54 & +44 & 05 & 12 & 0.198 & 2 & 0.6810  & 0.55 \\
{\rm A~1878} & 14 & 12 & 52 & +29 & 14 & 28 & 0.220 & 2 &  0.7894 & 0.70  \\
{\rm A~2111} & 15 & 39 & 40 & +34 & 25 & 27 & 0.229 & 2 & 0.8030  & 0.70 \\
{\rm A~1952} & 14 & 41 & 03 & +28 & 37 & 00 & 0.248 & 2 & 0.7989  & 0.55-0.80\\
\hline
         \end{array}
      \]
\label{tab:all}
   \end{table*}


\section{Galaxy detection and extraction}

We used SExtractor \citep{bertin96} in order to detect the individual objects in our images, and to extract their photometric parameters. The extraction of the galaxies was performed on the deeper Gunn-r images. The photometry of the galaxies in the B-band was obtained using the ASSOC mode of SExtractor. We fixed several SExtractor parameters based on the properties of the images, such as: background level, stellar FWHM, zero-point, exposure time, or radius for aperture photometry. The values of the other SExtractor parameters such as the minimum area of pixels above threshold (DETECT\_MINAREA), the number of deblending sub-thresholds (DEBLEND\_NTHRESH) or the minimum contrast parameter for deblending (DEBLEND\_MINCONT), were established after checking the results of the deblending images.


In this way, we obtained a first catalogue of 488 objects, including stars and galaxies. The stars were identified using the stellar index given by SExtractor. We considered an object as a galaxy when its stellar index was smaller than 0.2 and as star if the stellar index was larger than 0.8. The objects with intermediate values of the stellarity index were considered as doubtful objects. Only a small fraction of objects were classified as stars (27) or doubtful (5) so the final catalogue includes 456 galaxies. 

SExtractor provides different magnitudes for each detected galaxy. We have considered two of them. The first one corresponds to a fixed-aperture of radius five kpc, useful to compare colors in the same physical region, \cite{varela04Tes}. The other one is the magnitude called by SExtractor 'MAG$\_$BEST' that is determined in an automatic aperture which depends on the neighbours around the galaxy. If those neighbours are bright enough to affect the magnitude corresponding to an aperture enclosing the whole object by more than 10\%, then that magnitude is taken as the corrected isophotal magnitude, which corresponds to the isophotal magnitude together with a correction. This magnitude provides the best measures of the total light of the objects \citep{nelson02,stott08}.  

The B- and r- SExtractor magnitudes were k-corrected in the following way. For the B-band filter we adopted the relation $k_B = 4.4225 z+0.0294$, obtained as an interpolation of the data given by \cite{pence76} in the range 0.08 $ \leq  z  \leq$ 0.24. The magnitudes of the Gunn-r filter were k-corrected using the approximation $k_r = 2.5  log (1+z)$, (\cite{jorgensen92})

In Table~\ref{tab:allerr} we show the mean r-band magnitude error provided by SExtractor for all the galaxies in each cluster. The last column shows the errors in colour, obtained as the quadratic sum of the errors of the fixed-aperture magnitude in the two filters B and r. For the calibration errors in the Gunn-r band, see \cite{fasano02}

\begin{table*}[h]º
      \caption[]{Photometric errors}
      \[
         \begin{array}{llll}
            \hline\noalign{\smallskip}
\multicolumn{1}{c}{\rm Name}&
\multicolumn{1}{c}{\rm Err \, Aper}&
\multicolumn{1}{c}{\rm Err \, Best}&
\multicolumn{1}{c}{\rm Err \, Col }\\
\hline\noalign{\smallskip}
{\rm A~2658} & 0.007 & 0.008 & 0.028\\
{\rm A~1643} & 0.005 & 0.006 & 0.033 \\
{\rm A~1878} &0.007 & 0.008 & 0.052 \\
{\rm A~2111} & 0.007 & 0.009 & 0.045\\
{\rm A~1952} & 0.006 & 0.007 & 0.040\\
\hline
         \end{array}
      \]
\label{tab:allerr}
   \end{table*}

\section{Properties of the cluster galaxies}

\subsection{The Color-Magnitude Relation}

In the early fifties, \cite{baum59} and \cite{rood69} realized that the color index of the bright, early type galaxies correlates with their luminosity. Later on, \cite{visvanathan77b} and \cite{visvanathan77}, concluded on the universality of the so called CMR found for early type galaxies. Since then the study of the CMR has been used as a tool to analyze the evolution with z of the early type galaxies (see for example \cite{driver06,deLucia07a}). 

\begin{figure}[h]
\centering
\includegraphics[clip,width=1.\hsize]{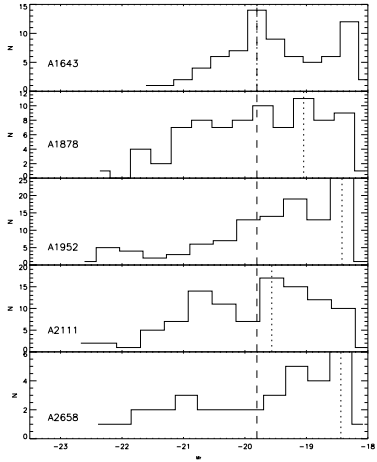} 
\caption{Absolute magnitude histogram of the galaxies in the five clusters. The dotted line shows the completeness magnitude limit for each cluster, whereas the dashed line shows the common magnitude limit we have adopted for the five clusters.}
\label{fig:mags}
\end{figure}

We present here the CMR for the galaxies found in our clusters. We use the color index B-r measured in a five kpc aperture. For the magnitudes we use the BEST magnitudes provided by SExtractor. All the detected galaxies in each cluster up to some limit magnitude were used to build the relation. In Figure~\ref{fig:mags}, we have plotted the absolute magnitude distribution of the sample together with the completeness limit for each cluster and for the whole sample respectively. The completeness limit has been defined as the maximum of the distribution. The sample appears to be complete up to M$_r \approx$ -19.8. Therefore, to avoid possible problems with the magnitude limit, we have considered a safer limit. Only  galaxies brighter than M$_r$=-20 have been used for the analysis of the CMR.

The definitive criterion to find the galaxies that actually belong to a given cluster is indeed the redshift. Unfortunately, the redshift information is in general scant for clusters at redshift $\sim$ 0.2 except for some particular cases. We have found in the literature 22 galaxies in the frame of A2111 with redshift data provided by \cite{miller06}, whereas for the other clusters there are just one or two redshift entries in the NED. On the other hand, the CMR is well defined and it is not necessary to have the redshift information at hand to analyze it. Background galaxies are identified as those objects that are  0.2 magnitudes redder than the value from the fitted CMR. That value seems safe if we take into account the combined uncertainty of the photometric errors and that 3$\sigma \approx$ 0.1-0.18. After applying this criterion the final number of galaxies retained as members of one of our five clusters amounts to 408. They are collected in the table presented in the appendix. The first column of that table gives the name of the cluster. The second and third columns give the coordinates of the galaxy, whereas we give in the fourth column the z information when available. The fifth and sixth columns give the r and B absolute magnitudes of each galaxy, assuming that they are located at the cluster redshift.

\begin{figure}[h]
\centering
\includegraphics[clip,width=1.\hsize]{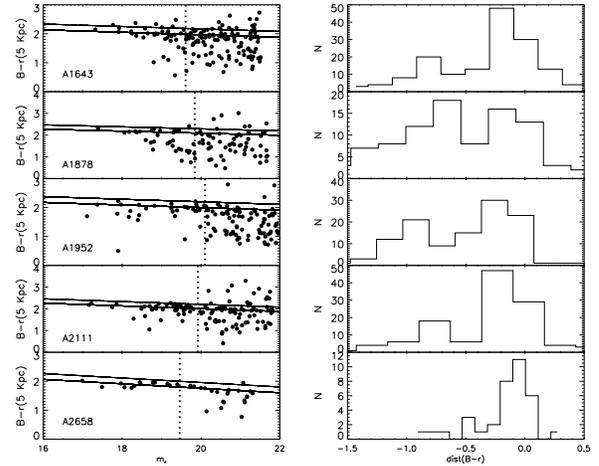} 
\caption{Left panels: The color-magnitude diagrams for the five clusters.  The solid line is the fit to the red sequence and the dotted line is the upper 0.2 magnitude limit. The vertical line corresponds to the limit M$_r$ = -20 at the cluster redshift. Right panels: The histograms of the B-r distance of the galaxies to the corresponding red sequence}
\label{fig:CM}
\end{figure}

The fit of the red sequence of the CMR for each cluster has been determined by carrying out a least absolute deviation regression fit to the observed data \citep{armstrong78}. The fit of the CMR for each cluster was obtained using an iterative procedure. A first fit was obtained using all the galaxies brighter than $M_r=-20$ for a given cluster. Then, the distance of each galaxy in B-r to the fitted CMR was computed. Those galaxies with a distance larger than three times the $rms$ of the fitted relation were rejected, and a new fit to the CMR was done with the remaining ones. This process was repeated until the fit to the CMR did not change anymore. The final fit has been estimated using a nonparametric bootstrap method, \cite{efron86}, with $n \log^2 n$ resamplings, being $n$ the number of galaxies up to the completeness limit, as prescribed in \cite{babu83}.  The slope and zero point are the median value of the resampling, while the standard errors have been estimated as the $rms$ of the bootstrap samples. In any case, we have checked that the early type galaxies that belong to A2111 are contained in the CMR. In Figure~\ref{fig:CM}, we show the colour-magnitude diagrams for all the galaxies in the frames and the CMRs fitted for all the clusters.  We have also plotted in that Figure the histogram of the color differences between the observed and the CMR-fitted values. We give in Table~\ref{tab:CMfb} the zero point, $a_{0}$, the slope, $a_{1}$ and the $rms$ of the fitted CMRs for each cluster.

\begin{table*}[h]
      \caption[]{CMR parameters and the galaxy blue fractions}
      \[
         \begin{array}{lccccc}
            \hline\noalign{\smallskip}
\multicolumn{1}{c}{\rm Name}&
\multicolumn{1}{c}{a0}&
\multicolumn{1}{c}{a1}&
\multicolumn{1}{c}{rms}&
\multicolumn{1}{c}{f_b (420 kpc)}&
\multicolumn{1}{c}{f_b (735 kpc)} \\
\hline\noalign{\smallskip}
{\rm A~2658} &  3.301 \pm 0.257 & -0.077 \pm 0.013 & 0.037 & 0.083 \pm 0.079 & \\
{\rm A~1643} &  2.825 \pm 0.224 & -0.043 \pm 0.011 & 0.035 & 0.090 \pm 0.086 & 0.090 \pm 0.086 \\
{\rm A~1878} &  3.022 \pm 0.390 & -0.046 \pm 0.021 & 0.060 & 0.363 \pm 0.102 & 0.517 \pm 0.092  \\ 
{\rm A~2111} &  3.285 \pm 0.079 & -0.063 \pm 0.004 & 0.053 & 0.031 \pm 0.030 &0.125 \pm 0.052 \\ 
{\rm A~1952} &  2.893 \pm 0.257 & -0.044 \pm 0.013 & 0.009 & 0.250 \pm 0.088 & 0.285 \pm 0.085  \\
\hline
         \end{array}
      \]
\label{tab:CMfb}
   \end{table*}
   
\begin{figure}[h]
\centering
\includegraphics[clip,width=0.8\hsize]{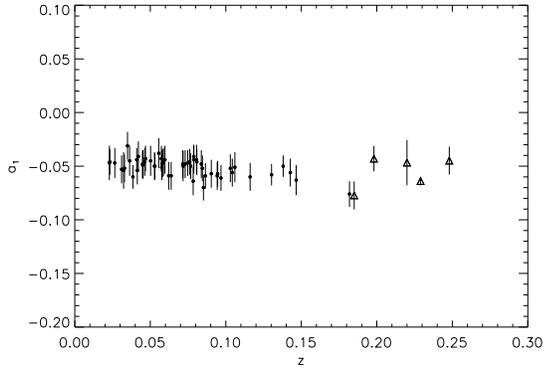}
\caption{Slopes of the CMR for the sample of \cite{lopezcruz04} (black circles) and our sample (empty triangles).}
\label{fig:slopesCM}
\end{figure}

In Figure~\ref{fig:slopesCM}, we have plotted the slopes of the fitted CMRs in our clusters at medium redshift together with those obtained by \cite{lopezcruz04} for clusters with $z < 0.15$. As the figure illustrates, there is no clear tendency of the slope of the CMR with redshift. The mean value of the slope of the CMR for our sample together with \cite{lopezcruz04} is $-0.051 \pm 0.008$, and only for our sample is  $-0.055 \pm 0.014$. Those values are also very similar to the slope value found by \cite{mei06} for two clusters at z$\sim$1.26. In other words, the slope values we find for our clusters at z $\sim$ 0.2 are completely consistent with the values found for lower and  much higher redshift values. Moreover, the range of values found at any redshift are also similar. Thus, we find no indication of a change of the CMR slope up to z $\sim$ 0.25 and even up to z$\sim$1.26. This result would indicate that the stellar population of the bright, early type galaxies defining the cluster red sequence was settled soon after the galaxy formation. 

\subsection{The blue galaxy fraction.}

We have studied the fraction of blue galaxies, f$_b$ in the bright population, M$_r \leq -20$,  of the clusters presented here. That blue fraction was defined as the ratio of the number of blue galaxies observed out of the total number of galaxies within a fixed aperture. We have considered blue galaxies those with B-r color at least 0.26 magnitudes bluer than the red sequence. This color index corresponds to the original definition by \cite{butcher84}, who considered as blue those galaxies whose colors were bluer than B-V = 0.2 at z=0. Given the photometric errors and the statistical nature of the k-correction we have just adopted that common value of the color index for all the five clusters in spite of their differences in redshift. The results are not affected if individual color values were adopted. 

The removal of foreground galaxies was done on the base of the measured redshift when available (just very few cases) or using statistical arguments. By integrating the luminosity function of field galaxies up to  M$_r=-20$ in the solid angle corresponding to each of our clusters, we obtained 0.38 foreground galaxies per frame at z = 0.2. This estimation is in good agreement with previous findings by  \cite{fasano00}. The foreground contamination is therefore statistically negligible. The fraction of blue galaxies has been computed for each cluster using all the surveyed area. In order to be able to compare our results for the different clusters and with other studies, we have considered that our frames are representative of the area corresponding to a circular aperture that, set at the center of the cluster, includes all the area that we have actually covered. For comparison purposes, we have adopted two apertures, of radius 420 and 735 kpc respectively. For the cluster A2658, only the smaller aperture could be used. Indeed, in the original definition given by \cite{butcher84}, the fraction was calculated for an aperture  containing 30\% of the cluster population (R$_{30}$). Since only the central parts of our clusters were sampled we could not determine the value of R$_{30}$ for them. The fixed apertures we have used are a substitute of the canonical value. We notice that they are in the range of the expected  R$_{30}$ values as given by \cite{butcher84}. The errors attributed to the measured fractions, listed in the last two columns of Table~\ref{tab:CMfb}, were computed assuming Poissonian statistics, following De Propris et al. (2004).

\begin{figure}[h]
\centering
\includegraphics[clip,width=0.8\hsize]{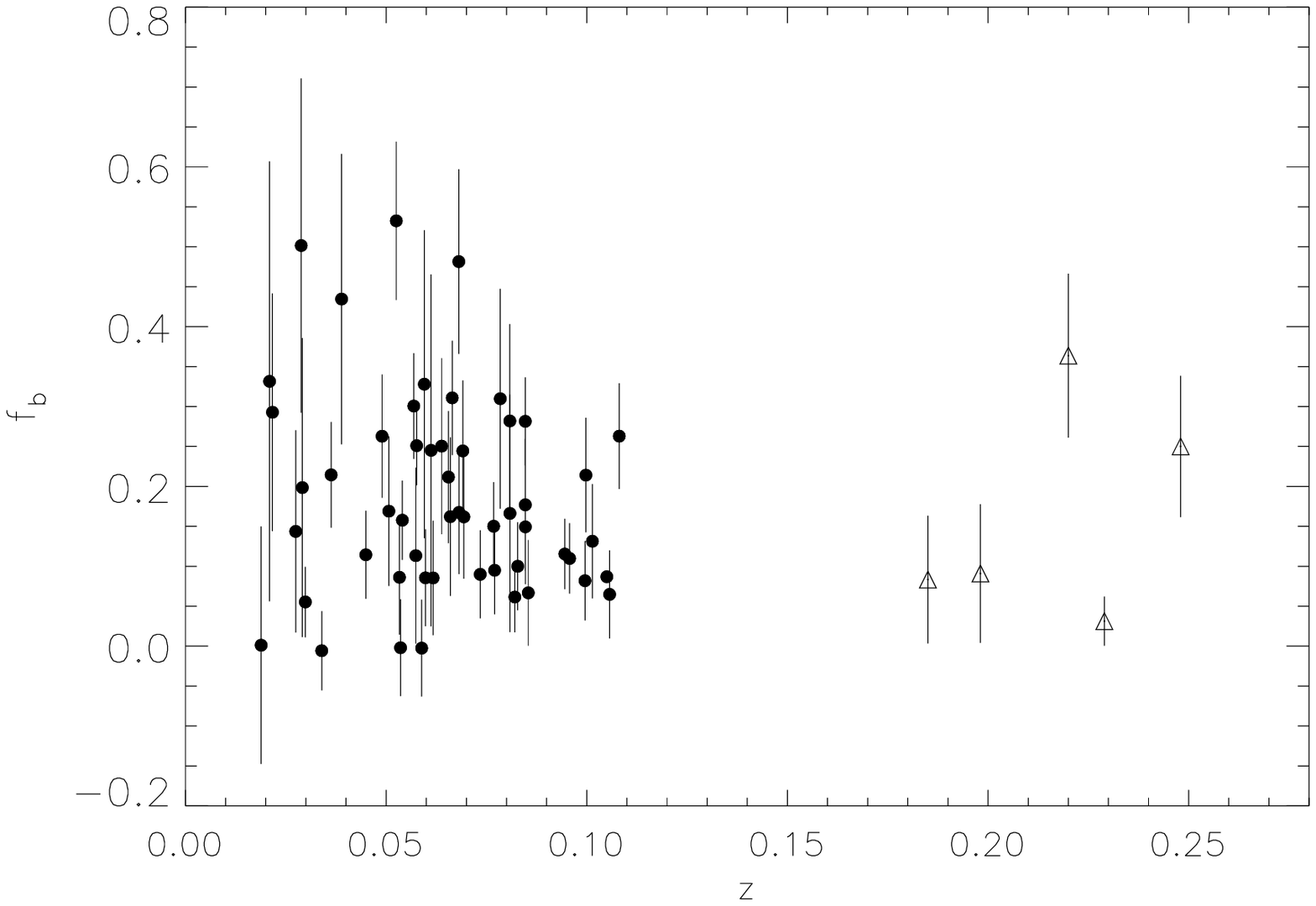}
\includegraphics[clip,width=0.8\hsize]{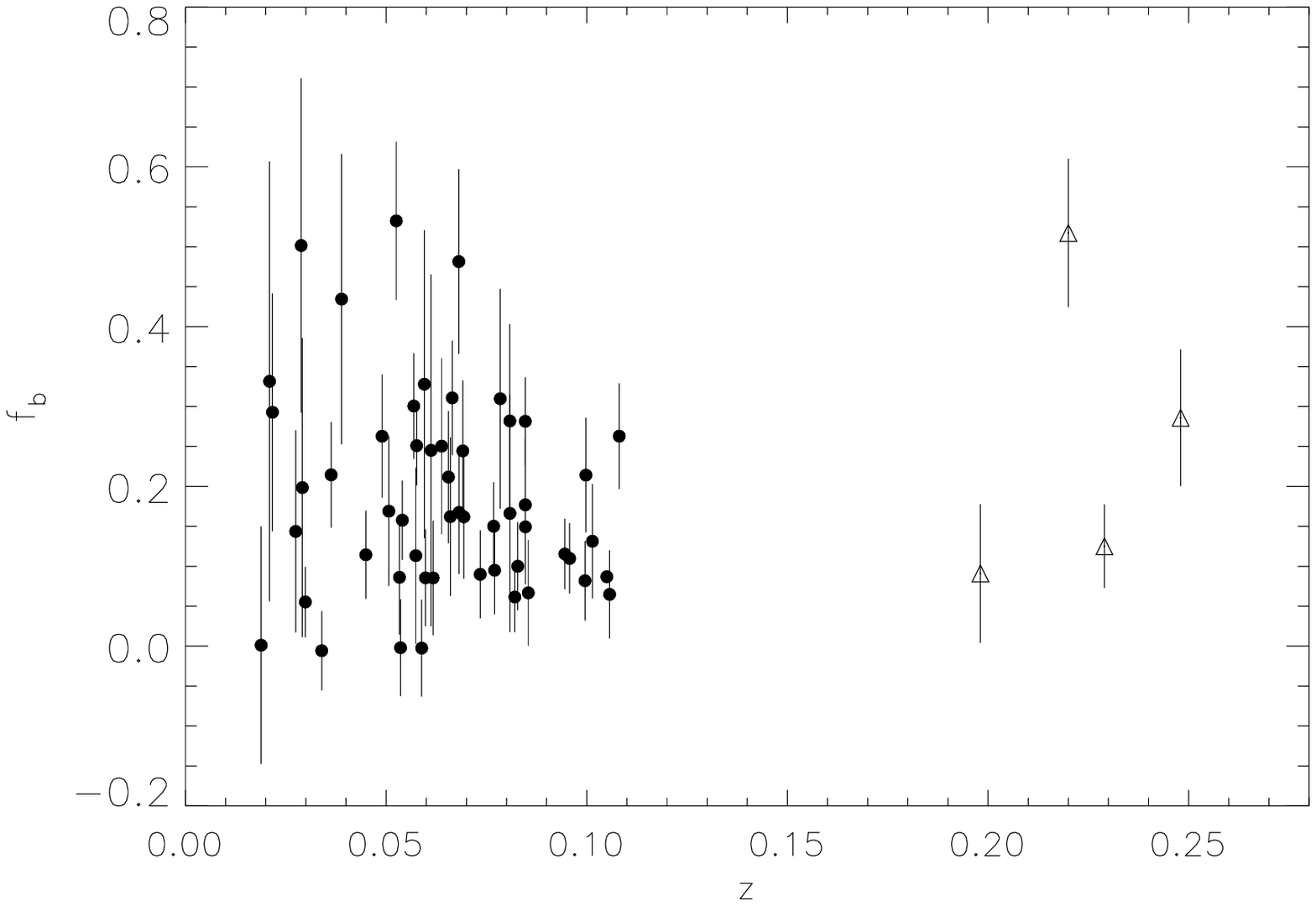}
\caption{Blue fraction of galaxies in our sample of clusters (triangles) compared with those obtained by \cite{depropris04} (black circles) for an aperture of $r_{200}/2$. The top (bottom) panel corresponds to the blue fraction computed within a radius of 420 kpc (735 kpc).}
\label{fig:BO}
\end{figure}

In Figure~\ref{fig:BO} we show the blue fraction of galaxies in our clusters as a function of redshift within a radius of 420 kpc (top panel) and within a radius of 735 kpc (bottom panel). We have also plotted for comparison the blue fraction of galaxies obtained from a sample of nearby galaxy clusters by \cite{depropris04} within an aperture of $r_{200}/2$. As can be seen in the Figure, our errors bars are very similar to those given by \cite{depropris04}. In all cases, we have more than 10 galaxies per cluster to compute the blue fraction. The comparison with the data by \cite{depropris04}, clearly indicates that there is no relation between the value of the blue galaxy fraction and the cluster redshift. 

The range of values found is also similar to that found by \cite{depropris04} for lower redshift clusters. In particular, the very high blue fraction we obtain for A1878 is found for some lower z clusters in the quoted reference. The central median values  we find for our sample are $<f_{b}>$=0.090 $\pm$ 0.138 for the 420 kpc and $0.285 \pm 0.194$ for the 735 kpc aperture, in agreement with the median $f_{b}$ value, 0.162 $\pm$ 0.125 of \cite{depropris04}, for an aperture of $r_{200}/2$. 

We find a nominal increment in the blue fraction as a function of the aperture. This is in agreement with the findings by \cite{margoniner00,goto03,depropris04}. 

Regarding the cluster A2111, \cite{butcher84} obtained a blue fraction of 0.16 $\pm$ 0.03 within a $r_{30}$ that, for this cluster, corresponds to 892 kpc. \cite{miller06} obtained, for the same aperture, the values of 0.15 $\pm$ 0.03 and 0.23 $\pm$ 0.03 using photometric data or only galaxies with spectroscopic data, respectively. We have obtained  0.031 $\pm$ 0.030  and 0.125 $\pm$ 0.052 for our 420 kpc and 735 kpc aperture, a smaller value, in agreement with the smaller aperture, even if not significantly different when the errors are taken into account.

\section{Galaxy Morphology}

For the purpose of the analysis presented here, we have classified the galaxies visually. A quantitative analysis will be presented in a forthcoming work, (Ascaso et al. 2008). 

\begin{figure}[h]
\centering
\includegraphics[clip,width=0.8\hsize]{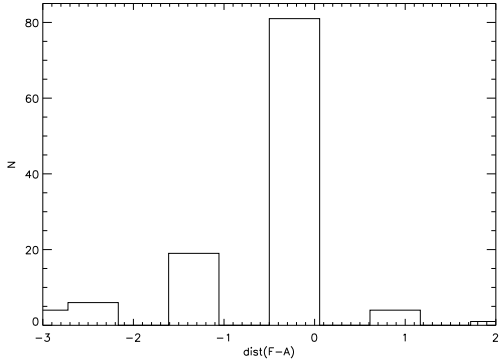}
\caption{The visual classification differences between \cite{fasano00} and this paper.}
\label{fig:morFY}
\end{figure}

All the galaxies brighter than $M_{r}=-20.0$ were classified visually by two of us (B.A. and J.V.) into four different types: Ellipticals (E), Lenticulars (S0), Spirals (Sp) and Irregulars (I). We have compared our classification with that reported by \cite{fasano00} for the galaxies in common. In Figure~\ref{fig:morFY}, we show the result of that comparison. Notice that 70$\%$ of the galaxies were classified with the same type, whereas 20$\%$ more differ by only one type. The morphological classification for that bright subsample is given in the last column of the table in the appendix.

In Table~\ref{tab:MorphType}, we show the percentages of the different galaxy types in the central part of each cluster in the 420 and 735 kpc aperture respectively. Notice that A1643 and A1878 have a large number of late type galaxies (about 60\%). In particular, A1878 contains also  a large fraction of irregular galaxies (19\%). The rest of the clusters have early-type population dominating in their cores. A diversity is clear as far as morphological populations is concerned.

\begin{table*}[h]
      \caption[]{Fraction of Morphological Types}
      \[
         \begin{array}{ccccccccc}
            \hline\noalign{\smallskip}
\multicolumn{1}{c}{\rm Name}&
\multicolumn{1}{c}{}&
\multicolumn{1}{c}{\rm 420 kpc}&
\multicolumn{1}{c}{ }&
\multicolumn{1}{c}{ }&
\multicolumn{1}{c}{}&
\multicolumn{1}{c}{\rm 735 kpc}&
\multicolumn{1}{c}{ }&
\multicolumn{1}{c}{ }\\
\multicolumn{1}{c}{}&
\multicolumn{1}{c}{\rm E }&
\multicolumn{1}{c}{\rm S0 }&
\multicolumn{1}{c}{\rm S }&
\multicolumn{1}{c}{\rm I }&
\multicolumn{1}{c}{\rm E }&
\multicolumn{1}{c}{\rm S0 }&
\multicolumn{1}{c}{\rm S }&
\multicolumn{1}{c}{\rm I }\\
\hline\noalign{\smallskip}
{\rm A~2658} & 0.54 & 0.31 & 0.15 & 0.00 &   &  &  &   \\
{\rm A~1643} & 0.22 & 0.22 & 0.56 & 0.00 & 0.24 & 0.19 & 0.57 & 0.00 \\
{\rm A~1878} & 0.11 & 0.22 & 0.41 & 0.26 & 0.14 & 0.24 & 0.43 & 0.19 \\ 
{\rm A~2111} & 0.38 & 0.28 & 0.28 & 0.00 & 0.35 & 0.28 & 0.30 & 0.08 \\ 
{\rm A~1952} & 0.52 & 0.28 & 0.20 & 0.00 & 0.45 & 0.31 & 0.24 & 0.00 \\
\hline
         \end{array}
      \]
\label{tab:MorphType}
   \end{table*}
   
\subsection{Morphological Distribution of the galaxies in the clusters}

It is well known  that early-type galaxies in clusters at low redshift are in general located in denser regions and closer to the center of the cluster than later types, \citep{dressler80}. We want now to investigate the way that those clusters at medium redshift are populated.

\begin{figure}[h]
\centering
\includegraphics[clip,width=1.0\hsize]{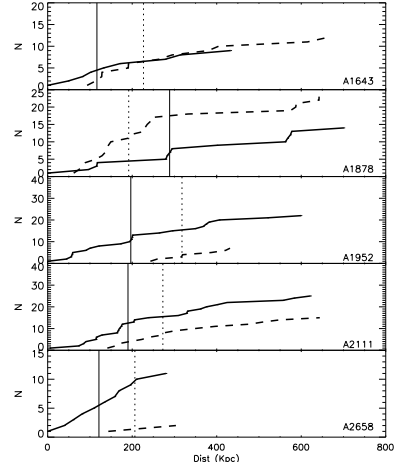}
\caption{Cumulative functions of the different morphological types as a function of the projected radius to the center of the cluster. Early types: solid lines; late types: dotted lines. The vertical lines indicate the radius where the distributions reach the 50\% level. }
\label{fig:typR}
\end{figure}

In Figure~\ref{fig:typR}, we have plotted the cumulative functions of the different types of galaxies versus projected distance of each galaxy to the center of the cluster. The solid lines represent the cumulative distribution of early-type, elliptical and lenticular, galaxies, whereas the dotted lines correspond to the cumulative distribution of  late-type galaxies, spiral and irregular. The vertical lines indicate the radius where the cumulative distributions reach the 50\% of the distributions. 

We see that all the clusters are dominated in their central regions by early type galaxies except A1878, that has a sizable fraction of late-type, including irregular galaxies. A fact that could explain its high (central) fraction of blue galaxies. This is however, not unique since similar cases can also be found at lower redshift, (see for example \cite{varela04Tes}). A1643 has also a large global fraction of late-type, spiral galaxies, but they do not dominate the core of the cluster. The rest of the clusters are also centrally dominated by a population of elliptical galaxies, with an overall population with a smaller fraction of late-type galaxies. As we noticed before, diversity seems to be the dominant aspect of our five clusters.

To test whether the distribution of early and late type galaxies are similar, we have performed a Kolmogorov-Smirnov test. Excluding A2658, for which there are not enough points to extract significant results, we find that the two populations are significantly different in two clusters, A1878 and A1952, while for the other two, A1643 and A2111, the KS test does not allow to extract significant results. Then, A1878 and A1952 show morphological segregation as found in low redshift samples as, for example, by \cite{adami98}. Moreover, in the case of A1878 the dominant morphological population are late type galaxies.

\subsection{The concentration parameter}

\begin{figure}[h]
\centering
\includegraphics[clip,width=1.0\hsize]{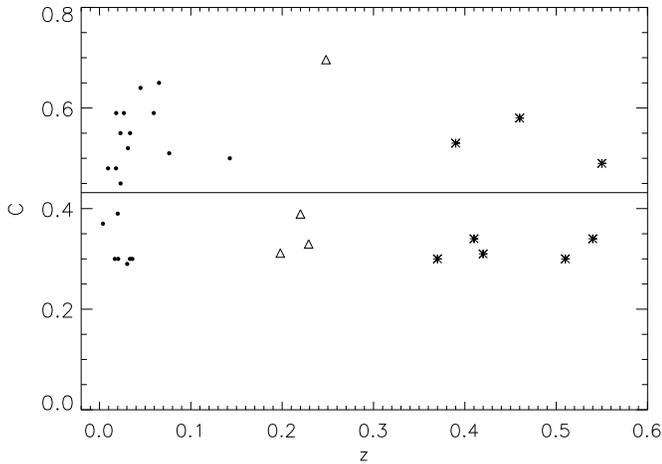}
\caption{Concentration parameter versus redshift for our clusters  (triangles), a low-redshift compilation (\cite{butcher78}: black points) and a higher redshift sample (\cite{dressler97}: asteriks). The horizontal line is the mean concentration value of the clusters with enough area coverage}
\label{fig:conD}
\end{figure}

We have calculated the concentration parameter of our clusters in the central 735 kpc.  We have worked it out using the  definition given by \cite{butcher78}, $\log$(R$_{60}$/R$_{20}$), where R$_{60}$ and R$_{20}$ are the radii containing 60\% and 20\% of the cluster populations within 735 kpc. Only the four clusters with enough area coverage were analyzed. The concentration values we have found are 0.311 for A1643, 0.389 for A1878, 0.329 for A2111 and 0.696 for A1952) as can be seen in the Figure~\ref{fig:conD}. We have overplotted the values found for lower z clusters by \cite{butcher78} and for a higher redshift sample by \cite{dressler97}. As can be seen in the figure, our concentration values span the full range of the values measured for lower redshfit clusters. Moreover, this range encompasses also that of the higher redshift clusters concentration values. It does not seem therefore, that there is any clear  tendency of the concentration parameter with redshift or morphological types.  At most, it could be argued that clusters tend to progressively populate the lower half of the plane when the redshift increases.

Likewise, \cite{butcher78,dressler97} suggested that the more irregular, less concentrated clusters would be preferentially populated by late type galaxies. In that sense, we notice that A1643, the cluster with the largest global fraction of late-type galaxies, presents the lowest value of the  concentration parameter.  Moreover, A1878, another cluster with a low concentration index presents also a rather high fraction of late type and irregular galaxies and, in fact, is dominated by this population. However, A2111, our third cluster with a low concentration, is dominated by an early-type population.  All in all, although there is  an indication for the higher fraction of irregular clusters with increasing redshift, the small statistics prevent us to extract a firm conclusion.

\subsection{Interaction systems}

\begin{figure}[h]
\centering
\includegraphics[clip,width=1.0\hsize]{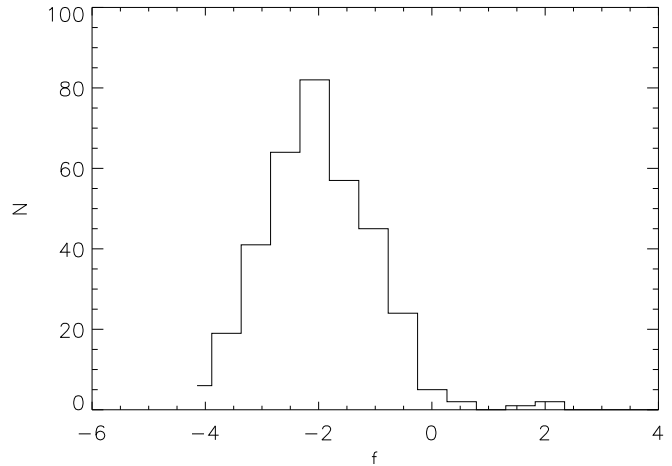}
\caption{Histogram of the f-parameter values for  the galaxies belonging to our clusters.}
\label{fig:histF}
\end{figure}

Other interesting feature that could deserve consideration in clusters at that range of redshift is the proportion of interacting systems compared to lower redshift clusters. To do that, we have calculated the distribution of the f-parameter defined by \cite{varela04} for the galaxies in the final catalogue. It gives an account of the relative importance of the tidal forces for every galaxy. Larger values indicate a larger proportion of interaction systems. The result is plotted in Figure~\ref{fig:histF}. The median value of the distribution is -1.59, whereas the median value found for the Coma Cluster amounts to -2.7 (\cite{varela04}). Moreover, we find that 63.97\% of the galaxies have a perturbation parameter higher than -2, which is the value chosen by \cite{varela04} to select truly interacting systems. 

More specifically, the median f-values we find are -1.39 (A2658), -1.92 (A1643),-1.60 (A1878), -1.67 (A2111) and -1.29 (A1952),  all of them larger than $-2$. This is indicative of the presence of a higher population of interacting systems in our sample than in the Coma Cluster.
 
\section{Discussion and Conclusions}

It is a very well known fact that locally, the color-magnitude relation for early-type galaxies has a well-defined slope with a small scatter, (\cite{bower92}).  Results based on the HST data, \citep{ellis97,stanford98,mei06}, demonstrated the existence of a tight red sequence in clusters at redshift up to 1.26, comparable in scatter and slope to that observed in the Coma Cluster and low redshift clusters. In agreement with these results, we find no significant difference in the slope of the  CMR for our cluster sample and other samples at different redshift values. This reinforces the conclusion on the universality of the CMR in all the explored z-range.

Regarding the Butcher-Oemler effect, the values we obtain for the blue galaxy fraction are similar to those found by \cite{depropris04}  for lower redshift SDSS clusters. The only outstanding case in our sample is A1878, with a large blue fraction. This goes together with an outstanding large fraction of late type and irregular galaxies in its central region and a low value of the concentration parameter. However, this case is not unique, (see for example, \citep{depropris04,aguerri07}). The blue fraction values that we find indicate that there is no evolution in this parameter up to redshift 0.25 at least. In other words, the Butcher-Oemler effect does start to manifest at even higher redshift. This is in agreement with previous results obtained with other samples, (see \cite{andreon06}). That conclusion, together with the universality of the CMR suggests that the stellar population in cluster galaxies has not evolved significantly in the last 2.5 Gyr.

We have found differences in the fraction of blue galaxies with the aperture considered for its determination, in the sense indicated by  other reported results, \citep{margoniner00,goto03,depropris04}, and in agreement with other values found in the literature for the blue fraction of A2111. However, these differences are not significant.

The morphological segregation between early and late-type galaxies, observed in nearby clusters \citep{adami98,aguerri04}, has been shown to be present in two clusters from our medium reshift sample. However, the range of situations we find is similar to that found at lower redshift and no evolutionary trend is manifested through it. 
 
We have also looked at the degree of interaction of the galaxies in our sample. We have found that the median value of the f-parameter is significantly higher than in the Coma cluster in all our five clusters. A wider sample at low redshift is lacking to extract conclusions about the existence of the possible higher incidence of interactions at higher redshift. 

The results we have obtained, suggest that the evolutionary status of clusters at z$\sim$0.2 is not significantly different from that of clusters at lower redshift. Indeed, larger samples at that intermediate redshift range are needed to establish the trends with z indicated by the concentration parameter are real or only artifacts produced by the small size of the available samples.

On the other hand, some parameters like the CMR slope is remarkably constant with redshift, strongly supporting the view that the red sequence was well established soon after the formation of the red, bright cluster galaxies. Regarding the Butcher-Oemler effect our results reinforces the view that it starts to appear at rather high z-values, in any case larger than z$\sim$0.25. ÊAs far as the morphological mixing is concerned, the values span a wide range, similarly to other samples at lower or higher redshift, indicating that here also the diversity is the dominant aspect. 


\begin{acknowledgements}

We acknowledge the anonymous referee for useful comments and suggestions. Bego\~na Ascaso is supported by an I3P Individual Fellowship from the Consejo Superior de Investigaciones Cientificas at the Instituto de Astrof\'isica de Andaluc\'ia. JALA acknowledges finalcial support by the project ESTALLIDOS   AYA2007-67965-C03-01.  We also acknowledge a grant by the Spanish Ministerio de Educacio«n y Ciencia, reference PNAYA2006-14056 and Proyecto de Excelencia of the Junta de Andalucia, reference J.A FQM 1392. 
\end{acknowledgements}

\clearpage
\onecolumn

\begin{appendix}
\section{Catalogue of galaxies detected in the cluster sample}
\begin{longtable}{lcccccccccc}
            \hline
\multicolumn{1}{c}{\rm Name}&
\multicolumn{3}{c}{$\alpha$ (J2000)}&
\multicolumn{3}{c}{$\delta$ (J2000)}&
\multicolumn{1}{c}{z}&
\multicolumn{1}{c}{M$_{r}$}&
\multicolumn{1}{c}{M$_{B}$}&
\multicolumn{1}{c}{Morph}\\
\hline\noalign{\smallskip}
{\rm A~2658} &  23h & 44m  & 47.99s &  -12d & 18m  & 46.20s &             &  -19.12  &  -18.24  &	    \\
{\rm A~2658} &  23h & 44m  & 55.21s &  -12d & 18m  & 37.00s &             &  -18.98  &  -18.20  &	    \\
{\rm A~2658} &  23h & 44m  & 49.55s &  -12d & 18m  & 34.40s &             &  -19.65  &  -18.99  &	    \\
{\rm A~2658} &  23h & 44m  & 49.62s &  -12d & 18m  & 31.90s &             &  -18.68  &  -18.01  &	    \\
{\rm A~2658} &  23h & 44m  & 50.35s &  -12d & 18m  & 25.50s &             &  -21.89  &  -21.06  &    S      \\
{\rm A~2658} &  23h & 44m  & 49.13s &  -12d & 18m  & 19.80s &             &  -19.29  &  -18.34  &	    \\
{\rm A~2658} &  23h & 44m  & 47.27s &  -12d & 18m  & 13.40s &             &  -19.24  &  -19.14  &	    \\
{\rm A~2658} &  23h & 44m  & 46.97s &  -12d & 18m  & 10.40s &             &  -20.94  &  -20.12  &   S0      \\
{\rm A~2658} &  23h & 44m  & 49.36s &  -12d & 18m  & 07.90s &             &  -18.44  &  -17.53  &	    \\
{\rm A~2658} &  23h & 44m  & 52.22s &  -12d & 18m  & 04.60s &             &  -19.99  &  -18.90  &    E      \\
{\rm A~2658} &  23h & 44m  & 54.99s &  -12d & 18m  & 05.60s &             &  -18.08  &  -17.23  &	    \\
{\rm A~2658} &  23h & 44m  & 54.27s &  -12d & 17m  & 59.30s &             &  -21.42  &  -20.39  &    E      \\
{\rm A~2658} &  23h & 44m  & 50.42s &  -12d & 17m  & 56.40s &             &  -19.39  &  -18.41  &	    \\
{\rm A~2658} &  23h & 44m  & 51.64s &  -12d & 17m  & 53.30s &             &  -18.71  &  -18.11  &	    \\
{\rm A~2658} &  23h & 44m  & 47.44s &  -12d & 17m  & 47.40s &             &  -20.92  &  -19.87  &    E      \\
{\rm A~2658} &  23h & 44m  & 50.34s &  -12d & 17m  & 32.70s &             &  -18.84  &  -20.30  &	    \\
{\rm A~2658} &  23h & 44m  & 49.28s &  -12d & 17m  & 38.20s &             &  -18.19  &  -19.65  &	    \\
{\rm A~2658} &  23h & 44m  & 50.26s &  -12d & 17m  & 20.90s &             &  -21.07  &  -20.35  &   S0      \\
{\rm A~2658} &  23h & 44m  & 49.84s &  -12d & 17m  & 26.70s &             &  -21.32  &  -20.95  &    E      \\
{\rm A~2658} &  23h & 44m  & 49.80s &  -12d & 17m  & 39.50s &             &  -22.39  &  -22.02  &    E      \\
{\rm A~2658} &  23h & 44m  & 54.96s &  -12d & 17m  & 38.80s &             &  -18.80  &  -18.35  &	    \\
{\rm A~2658} &  23h & 44m  & 55.87s &  -12d & 17m  & 37.60s &             &  -18.22  &  -17.40  &	    \\
{\rm A~2658} &  23h & 44m  & 51.86s &  -12d & 17m  & 35.30s &             &  -19.57  &  -18.55  &	    \\
{\rm A~2658} &  23h & 44m  & 47.85s &  -12d & 17m  & 31.10s &             &  -18.14  &  -17.23  &	    \\
{\rm A~2658} &  23h & 44m  & 50.96s &  -12d & 17m  & 19.10s &             &  -20.24  &  -19.17  &    E      \\
{\rm A~2658} &  23h & 44m  & 55.84s &  -12d & 17m  & 17.60s &             &  -20.18  &  -19.15  &   S0      \\
{\rm A~2658} &  23h & 44m  & 51.40s &  -12d & 17m  & 11.00s &             &  -18.35  &  -18.28  &	    \\
{\rm A~2658} &  23h & 44m  & 56.18s &  -12d & 17m  & 07.50s &             &  -21.14  &  -20.31  &    S      \\
{\rm A~2658} &  23h & 44m  & 51.13s &  -12d & 16m  & 48.00s &             &  -20.61  &  -19.49  &    E      \\
{\rm A~2658} &  23h & 44m  & 46.13s &  -12d & 16m  & 49.10s &             &  -18.82  &  -18.35  &	    \\
{\rm A~2658} &  23h & 44m  & 49.65s &  -12d & 16m  & 35.80s &             &  -20.56  &  -19.43  &   S0      \\
{\rm A~2658} &  23h & 44m  & 47.99s &  -12d & 16m  & 36.20s &             &  -18.29  &  -17.66  &	    \\
{\rm A~2658} &  23h & 44m  & 51.63s &  -12d & 16m  & 28.80s &             &  -18.70  &  -17.79  &	    \\
{\rm A~2658} &  23h & 44m  & 53.27s &  -12d & 16m  & 23.60s &             &  -18.27  &  -17.77  &	    \\
{\rm A~1643} &  12h & 55m  & 52.30s &	44d & 05m  & 47.30s &             &  -19.88  &  -20.46  &    S      \\
{\rm A~1643} &  12h & 55m  & 52.44s &	44d & 05m  & 52.70s &             &  -19.52  &  -20.82  &	    \\
{\rm A~1643} &  12h & 55m  & 54.14s &	44d & 05m  & 52.70s &             &  -19.79  &  -18.64  &	    \\
{\rm A~1643} &  12h & 55m  & 59.31s &	44d & 05m  & 53.20s &             &  -19.36  &  -17.98  &	    \\
{\rm A~1643} &  12h & 55m  & 53.80s &	44d & 03m  & 15.20s &             &  -19.79  &  -18.61  &    S      \\
{\rm A~1643} &  12h & 55m  & 55.18s &	44d & 03m  & 47.50s &             &  -20.67  &  -19.46  &   S0      \\
{\rm A~1643} &  12h & 55m  & 49.83s &	44d & 04m  & 08.80s &             &  -19.82  &  -19.24  &	    \\
{\rm A~1643} &  12h & 55m  & 49.75s &	44d & 04m  & 05.50s &             &  -20.61  &  -19.63  &    S      \\
{\rm A~1643} &  12h & 55m  & 47.93s &	44d & 04m  & 01.20s &             &  -20.36  &  -19.10  &    E      \\
{\rm A~1643} &  12h & 55m  & 48.06s &	44d & 04m  & 06.70s &             &  -18.24  &  -17.09  &	    \\
{\rm A~1643} &  12h & 55m  & 51.98s &	44d & 04m  & 05.90s &             &  -18.93  &  -19.06  &	    \\
{\rm A~1643} &  12h & 55m  & 59.67s &	44d & 04m  & 05.20s &             &  -19.64  &  -18.62  &    S      \\
{\rm A~1643} &  12h & 55m  & 53.06s &	44d & 04m  & 06.60s &             &  -18.89  &  -17.82  &	    \\
{\rm A~1643} &  12h & 55m  & 55.75s &	44d & 04m  & 07.30s &             &  -19.39  &  -18.88  &	    \\
{\rm A~1643} &  12h & 56m  & 01.43s &	44d & 04m  & 07.90s &             &  -19.71  &  -19.19  &	    \\
{\rm A~1643} &  12h & 55m  & 53.64s &	44d & 04m  & 13.70s &             &  -20.09  &  -19.36  &    S      \\
{\rm A~1643} &  12h & 55m  & 50.96s &	44d & 04m  & 31.00s &             &  -21.15  &  -20.05  &    E      \\
{\rm A~1643} &  12h & 55m  & 59.04s &	44d & 04m  & 26.90s &             &  -19.00  &  -17.81  &	    \\
{\rm A~1643} &  12h & 55m  & 55.35s &	44d & 04m  & 34.40s &             &  -20.69  &  -20.35  &    E      \\
{\rm A~1643} &  12h & 55m  & 54.88s &	44d & 04m  & 33.90s &             &  -20.14  &  -19.56  &    S      \\
{\rm A~1643} &  12h & 55m  & 56.61s &	44d & 04m  & 38.20s &             &  -18.70  &  -20.14  &	    \\
{\rm A~1643} &  12h & 55m  & 52.33s &	44d & 04m  & 46.80s &             &  -18.42  &  -19.14  &	    \\
{\rm A~1643} &  12h & 55m  & 52.97s &	44d & 04m  & 50.00s &             &  -19.82  &  -19.09  &	    \\
{\rm A~1643} &  12h & 55m  & 52.96s &	44d & 04m  & 39.20s &  0.1978     &  -19.73  &  -19.60  &    S      \\
{\rm A~1643} &  12h & 55m  & 52.70s &	44d & 04m  & 44.50s &             &  -20.42  &  -19.74  &   S0      \\
{\rm A~1643} &  12h & 55m  & 54.94s &	44d & 04m  & 45.60s &             &  -19.29  &  -19.15  &	    \\
{\rm A~1643} &  12h & 55m  & 48.16s &	44d & 04m  & 49.50s &             &  -19.49  &  -18.74  &	    \\
{\rm A~1643} &  12h & 55m  & 47.94s &	44d & 04m  & 51.60s &             &  -19.59  &  -18.95  &	    \\
{\rm A~1643} &  12h & 55m  & 51.98s &	44d & 04m  & 53.10s &             &  -19.57  &  -19.08  &	    \\
{\rm A~1643} &  12h & 55m  & 55.21s &	44d & 04m  & 53.10s &             &  -18.00  &  -17.50  &	    \\
{\rm A~1643} &  12h & 55m  & 54.40s &	44d & 04m  & 53.70s &             &  -18.26  &  -17.50  &	    \\
{\rm A~1643} &  12h & 55m  & 59.29s &	44d & 04m  & 57.10s &             &  -20.02  &  -18.94  &    S      \\
{\rm A~1643} &  12h & 55m  & 56.07s &	44d & 04m  & 58.00s &             &  -18.49  &  -17.59  &	    \\
{\rm A~1643} &  12h & 55m  & 54.00s &	44d & 05m  & 12.40s &             &  -21.61  &  -20.35  &   S0      \\
{\rm A~1643} &  12h & 56m  & 01.63s &	44d & 05m  & 09.10s &             &  -19.45  &  -18.11  &	    \\
{\rm A~1643} &  12h & 55m  & 49.61s &	44d & 05m  & 09.50s &             &  -18.25  &  -17.15  &	    \\
{\rm A~1643} &  12h & 55m  & 47.67s &	44d & 05m  & 15.70s &             &  -18.15  &  -17.15  &	    \\
{\rm A~1643} &  12h & 55m  & 54.61s &	44d & 05m  & 21.40s &             &  -19.23  &  -17.99  &	    \\
{\rm A~1643} &  12h & 55m  & 53.05s &	44d & 05m  & 23.40s &             &  -20.19  &  -18.97  &   S0      \\
{\rm A~1643} &  12h & 56m  & 00.41s &	44d & 05m  & 29.90s &             &  -19.20  &  -18.91  &	    \\
{\rm A~1643} &  12h & 55m  & 48.02s &	44d & 05m  & 35.90s &             &  -19.73  &  -18.91  &	    \\
{\rm A~1643} &  12h & 55m  & 52.36s &	44d & 05m  & 38.40s &             &  -19.16  &  -19.35  &	    \\
{\rm A~1643} &  12h & 55m  & 52.76s &	44d & 05m  & 37.90s &             &  -19.88  &  -19.97  &    S      \\
{\rm A~1643} &  12h & 55m  & 54.21s &	44d & 05m  & 44.70s &             &  -19.41  &  -18.29  &	    \\
{\rm A~1643} &  12h & 56m  & 01.53s &	44d & 03m  & 29.90s &             &  -18.65  &  -20.27  &	    \\
{\rm A~1643} &  12h & 55m  & 50.27s &	44d & 03m  & 30.50s &             &  -19.27  &  -18.68  &	    \\
{\rm A~1643} &  12h & 55m  & 53.07s &	44d & 05m  & 47.80s &             &  -18.67  &  -17.64  &	    \\
{\rm A~1643} &  12h & 55m  & 48.08s &	44d & 05m  & 51.70s &             &  -18.23  &  -17.80  &	    \\
{\rm A~1643} &  12h & 55m  & 34.43s &	44d & 08m  & 50.30s &             &  -19.53  &  -18.37  &	    \\
{\rm A~1643} &  12h & 55m  & 44.49s &	44d & 08m  & 53.60s &             &  -19.09  &  -18.13  &	    \\
{\rm A~1643} &  12h & 55m  & 45.49s &	44d & 06m  & 39.60s &             &  -18.15  &  -19.35  &	    \\
{\rm A~1643} &  12h & 55m  & 44.70s &	44d & 06m  & 35.60s &             &  -19.65  &  -18.94  &	    \\
{\rm A~1643} &  12h & 55m  & 38.43s &	44d & 06m  & 29.90s &             &  -18.50  &  -17.72  &	    \\
{\rm A~1643} &  12h & 55m  & 38.94s &	44d & 06m  & 35.20s &             &  -18.31  &  -17.92  &	    \\
{\rm A~1643} &  12h & 55m  & 45.18s &	44d & 06m  & 46.30s &             &  -19.68  &  -18.40  &    E      \\
{\rm A~1643} &  12h & 55m  & 32.98s &	44d & 06m  & 50.40s &             &  -19.92  &  -19.46  &    S      \\
{\rm A~1643} &  12h & 55m  & 33.62s &	44d & 06m  & 30.00s &             &  -18.06  &  -17.72  &	    \\
{\rm A~1643} &  12h & 55m  & 37.87s &	44d & 06m  & 57.10s &             &  -18.59  &  -17.40  &	    \\
{\rm A~1643} &  12h & 55m  & 46.43s &	44d & 06m  & 58.80s &             &  -18.14  &  -17.06  &	    \\
{\rm A~1643} &  12h & 55m  & 33.82s &	44d & 07m  & 12.50s &             &  -20.93  &  -19.67  &    E      \\
{\rm A~1643} &  12h & 55m  & 36.30s &	44d & 07m  & 15.70s &             &  -18.80  &  -18.47  &	    \\
{\rm A~1643} &  12h & 55m  & 41.25s &	44d & 07m  & 15.00s &             &  -18.32  &  -17.59  &	    \\
{\rm A~1643} &  12h & 55m  & 39.33s &	44d & 07m  & 21.30s &             &  -19.72  &  -18.39  &    S      \\
{\rm A~1643} &  12h & 55m  & 37.74s &	44d & 07m  & 23.30s &             &  -18.00  &  -17.63  &	    \\
{\rm A~1643} &  12h & 55m  & 38.60s &	44d & 07m  & 29.10s &             &  -18.09  &  -16.93  &	    \\
{\rm A~1643} &  12h & 55m  & 46.75s &	44d & 07m  & 35.40s &             &  -18.99  &  -17.90  &	    \\
{\rm A~1643} &  12h & 55m  & 42.78s &	44d & 07m  & 48.60s &             &  -18.24  &  -16.93  &	    \\
{\rm A~1643} &  12h & 55m  & 36.40s &	44d & 07m  & 53.40s &             &  -20.71  &  -20.57  &    I      \\
{\rm A~1643} &  12h & 55m  & 36.55s &	44d & 07m  & 54.10s &             &  -20.34  &  -20.10  &	    \\
{\rm A~1643} &  12h & 55m  & 36.63s &	44d & 08m  & 20.30s &             &  -20.17  &  -20.49  &    I      \\
{\rm A~1643} &  12h & 55m  & 36.38s &	44d & 08m  & 24.40s &             &  -19.77  &  -19.13  &   S0      \\
{\rm A~1643} &  12h & 55m  & 36.57s &	44d & 08m  & 30.40s &             &  -20.27  &  -19.84  &    E     \\
{\rm A~1643} &  12h & 55m  & 43.31s &	44d & 08m  & 28.90s &             &  -18.20  &  -18.04  &	   \\
{\rm A~1643} &  12h & 55m  & 38.31s &	44d & 08m  & 38.70s &             &  -18.02  &  -17.76  &	   \\
{\rm A~1643} &  12h & 55m  & 37.59s &	44d & 06m  & 21.10s &             &  -19.42  &  -18.66  &    S     \\
{\rm A~1878} &  14h & 12m  & 54.12s &	29d & 16m  & 16.60s &             &  -18.74  &  -17.61  &	    \\
{\rm A~1878} &  14h & 12m  & 49.83s &	29d & 13m  & 40.60s &             &  -18.90  &  -18.43  &	    \\
{\rm A~1878} &  14h & 12m  & 47.43s &	29d & 13m  & 55.50s &             &  -18.53  &  -20.39  &	    \\
{\rm A~1878} &  14h & 12m  & 47.82s &	29d & 13m  & 53.40s &             &  -21.69  &  -20.68  &    S      \\
{\rm A~1878} &  14h & 12m  & 53.32s &	29d & 13m  & 47.00s &             &  -18.37  &  -18.29  &	    \\
{\rm A~1878} &  14h & 12m  & 54.23s &	29d & 13m  & 57.60s &             &  -20.23  &  -20.50  &    S      \\
{\rm A~1878} &  14h & 12m  & 50.11s &	29d & 13m  & 59.90s &             &  -18.63  &  -19.32  &	    \\
{\rm A~1878} &  14h & 12m  & 49.97s &	29d & 14m  & 02.60s &             &  -20.45  &  -19.30  &    S      \\
{\rm A~1878} &  14h & 12m  & 56.80s &	29d & 14m  & 03.60s &             &  -20.38  &  -20.06  &    I      \\
{\rm A~1878} &  14h & 12m  & 54.78s &	29d & 14m  & 03.90s &             &  -18.55  &  -17.39  &	    \\
{\rm A~1878} &  14h & 12m  & 47.17s &	29d & 14m  & 05.80s &             &  -20.04  &  -19.70  &    I      \\
{\rm A~1878} &  14h & 12m  & 49.47s &	29d & 14m  & 09.90s &             &  -21.57  &  -20.53  &    S      \\
{\rm A~1878} &  14h & 12m  & 49.03s &	29d & 14m  & 07.80s &             &  -18.94  &  -21.00  &	    \\
{\rm A~1878} &  14h & 12m  & 52.50s &	29d & 14m  & 11.40s &             &  -20.94  &  -20.28  &    S      \\
{\rm A~1878} &  14h & 12m  & 54.85s &	29d & 14m  & 17.30s &             &  -19.91  &  -19.67  &    S      \\
{\rm A~1878} &  14h & 12m  & 54.65s &	29d & 14m  & 23.80s &             &  -19.23  &  -19.19  &	    \\
{\rm A~1878} &  14h & 12m  & 47.85s &	29d & 14m  & 17.10s &             &  -19.70  &  -18.95  &	    \\
{\rm A~1878} &  14h & 12m  & 54.15s &	29d & 14m  & 19.30s &             &  -20.80  &  -19.49  &    E      \\
{\rm A~1878} &  14h & 12m  & 52.75s &	29d & 14m  & 20.20s &             &  -18.72  &  -20.18  &	    \\
{\rm A~1878} &  14h & 12m  & 52.18s &	29d & 14m  & 28.40s &  0.2220     &  -22.36  &  -21.69  &    E     \\
{\rm A~1878} &  14h & 12m  & 46.85s &	29d & 14m  & 26.40s &             &  -21.02  &  -20.50  &    I      \\
{\rm A~1878} &  14h & 12m  & 54.72s &	29d & 14m  & 31.90s &             &  -21.42  &  -20.23  &    E      \\
{\rm A~1878} &  14h & 12m  & 56.29s &	29d & 14m  & 31.40s &             &  -20.24  &  -19.80  &    I      \\
{\rm A~1878} &  14h & 12m  & 51.24s &	29d & 14m  & 48.20s &             &  -20.10  &  -20.01  &    S      \\
{\rm A~1878} &  14h & 12m  & 51.04s &	29d & 14m  & 39.30s &             &  -19.72  &  -20.22  &	    \\
{\rm A~1878} &  14h & 12m  & 50.98s &	29d & 14m  & 42.30s &             &  -20.84  &  -21.55  &    I      \\
{\rm A~1878} &  14h & 12m  & 46.74s &	29d & 14m  & 40.00s &             &  -18.44  &  -18.10  &	    \\
{\rm A~1878} &  14h & 12m  & 53.29s &	29d & 14m  & 41.40s &             &  -20.30  &  -20.22  &	    \\
{\rm A~1878} &  14h & 12m  & 53.32s &	29d & 14m  & 44.60s &             &  -19.55  &  -21.50  &	    \\
{\rm A~1878} &  14h & 12m  & 49.12s &	29d & 14m  & 42.50s &             &  -21.38  &  -20.33  &    S      \\
{\rm A~1878} &  14h & 12m  & 50.12s &	29d & 14m  & 47.30s &             &  -20.40  &  -19.13  &   S0      \\
{\rm A~1878} &  14h & 12m  & 52.25s &	29d & 14m  & 53.70s &             &  -20.41  &  -20.57  &    S      \\
{\rm A~1878} &  14h & 12m  & 51.99s &	29d & 14m  & 57.10s &             &  -19.53  &  -19.92  &	    \\
{\rm A~1878} &  14h & 12m  & 50.96s &	29d & 14m  & 56.60s &             &  -21.29  &  -20.33  &    S      \\
{\rm A~1878} &  14h & 12m  & 46.14s &	29d & 14m  & 55.60s &             &  -19.94  &  -19.35  &   S0      \\
{\rm A~1878} &  14h & 12m  & 46.58s &	29d & 14m  & 59.10s &             &  -20.94  &  -19.69  &   S0      \\
{\rm A~1878} &  14h & 12m  & 53.29s &	29d & 14m  & 56.90s &             &  -18.53  &  -17.55  &	    \\
{\rm A~1878} &  14h & 12m  & 48.23s &	29d & 15m  & 01.10s &             &  -19.28  &  -18.27  &	    \\
{\rm A~1878} &  14h & 12m  & 50.01s &	29d & 15m  & 05.00s &             &  -18.23  &  -17.10  &	    \\
{\rm A~1878} &  14h & 12m  & 56.61s &	29d & 15m  & 05.30s &             &  -19.18  &  -19.30  &	    \\
{\rm A~1878} &  14h & 12m  & 49.37s &	29d & 15m  & 12.10s &             &  -18.79  &  -17.72  &	    \\
{\rm A~1878} &  14h & 12m  & 50.60s &	29d & 15m  & 13.20s &             &  -18.64  &  -18.42  &	    \\
{\rm A~1878} &  14h & 12m  & 49.68s &	29d & 15m  & 14.20s &             &  -19.56  &  -19.22  &	    \\
{\rm A~1878} &  14h & 12m  & 55.12s &	29d & 15m  & 14.70s &             &  -18.85  &  -17.83  &	    \\
{\rm A~1878} &  14h & 12m  & 51.24s &	29d & 15m  & 22.10s &             &  -19.89  &  -19.95  &    S      \\
{\rm A~1878} &  14h & 12m  & 53.39s &	29d & 15m  & 22.30s &             &  -19.01  &  -18.65  &	    \\
{\rm A~1878} &  14h & 12m  & 51.04s &	29d & 15m  & 28.90s &             &  -19.51  &  -21.46  &	    \\
{\rm A~1878} &  14h & 12m  & 53.49s &	29d & 15m  & 27.70s &             &  -18.76  &  -18.14  &	    \\
{\rm A~1878} &  14h & 12m  & 52.60s &	29d & 15m  & 41.10s &             &  -18.61  &  -18.01  &	    \\
{\rm A~1878} &  14h & 12m  & 52.43s &	29d & 15m  & 48.70s &             &  -21.00  &  -19.79  &   S0      \\
{\rm A~1878} &  14h & 12m  & 55.73s &	29d & 15m  & 56.70s &             &  -18.18  &  -17.32  &	    \\
{\rm A~1878} &  14h & 12m  & 53.61s &	29d & 16m  & 00.80s &             &  -19.77  &  -18.60  &	    \\
{\rm A~1878} &  14h & 12m  & 53.04s &	29d & 16m  & 07.40s &             &  -18.86  &  -21.15  &	    \\
{\rm A~1878} &  14h & 12m  & 47.96s &	29d & 16m  & 09.50s &             &  -19.69  &  -19.16  &	    \\
{\rm A~1878} &  14h & 13m  & 00.54s &	29d & 13m  & 56.90s &             &  -21.15  &  -20.41  &   S0      \\
{\rm A~1878} &  14h & 12m  & 56.76s &	29d & 14m  & 03.60s &             &  -20.00  &  -20.14  &    I      \\
{\rm A~1878} &  14h & 12m  & 56.78s &	29d & 12m  & 00.30s &             &  -19.87  &  -18.86  &   S0      \\
{\rm A~1878} &  14h & 12m  & 57.80s &	29d & 12m  & 01.60s &             &  -20.47  &  -19.97  &   S0      \\
{\rm A~1878} &  14h & 12m  & 59.05s &	29d & 12m  & 14.40s &             &  -20.60  &  -20.01  &    E      \\
{\rm A~1878} &  14h & 12m  & 59.84s &	29d & 12m  & 19.50s &             &  -20.45  &  -21.96  &    S      \\
{\rm A~1878} &  14h & 13m  & 00.58s &	29d & 12m  & 22.90s &             &  -20.30  &  -19.90  &    S      \\
{\rm A~1878} &  14h & 13m  & 01.89s &	29d & 12m  & 17.50s &             &  -19.55  &  -18.93  &    S      \\
{\rm A~1878} &  14h & 13m  & 05.79s &	29d & 12m  & 20.80s &             &  -18.24  &  -18.48  &	    \\
{\rm A~1878} &  14h & 12m  & 58.97s &	29d & 12m  & 33.00s &             &  -18.13  &  -17.07  &	    \\
{\rm A~1878} &  14h & 13m  & 01.29s &	29d & 12m  & 36.90s &             &  -20.97  &  -20.65  &   S0      \\
{\rm A~1878} &  14h & 13m  & 05.52s &	29d & 12m  & 36.60s &             &  -18.59  &  -18.07  &	    \\
{\rm A~1878} &  14h & 13m  & 02.23s &	29d & 12m  & 40.50s &             &  -18.23  &  -18.12  &	    \\
{\rm A~1878} &  14h & 13m  & 05.38s &	29d & 12m  & 42.80s &             &  -18.24  &  -17.75  &	    \\
{\rm A~1878} &  14h & 12m  & 58.42s &	29d & 12m  & 53.60s &             &  -18.05  &  -19.13  &	    \\
{\rm A~1878} &  14h & 12m  & 58.26s &	29d & 12m  & 54.90s &             &  -19.49  &  -20.04  &	    \\
{\rm A~1878} &  14h & 13m  & 05.59s &	29d & 12m  & 54.20s &             &  -20.53  &  -19.81  &    E      \\
{\rm A~1878} &  14h & 13m  & 04.82s &	29d & 12m  & 55.40s &             &  -19.03  &  -18.41  &	    \\
{\rm A~1878} &  14h & 13m  & 02.81s &	29d & 12m  & 55.70s &             &  -19.44  &  -18.73  &    S      \\
{\rm A~1878} &  14h & 12m  & 58.70s &	29d & 12m  & 56.60s &             &  -18.92  &  -18.15  &	    \\
{\rm A~1878} &  14h & 13m  & 04.41s &	29d & 13m  & 00.70s &             &  -20.06  &  -19.78  &    S      \\
{\rm A~1878} &  14h & 12m  & 55.45s &	29d & 13m  & 04.30s &             &  -19.48  &  -19.63  &    I      \\
{\rm A~1878} &  14h & 12m  & 55.11s &	29d & 13m  & 09.90s &             &  -19.13  &  -19.64  &	    \\
{\rm A~1878} &  14h & 12m  & 57.07s &	29d & 13m  & 19.80s &             &  -18.08  &  -18.08  &	    \\
{\rm A~1878} &  14h & 12m  & 57.65s &	29d & 13m  & 22.20s &             &  -18.09  &  -17.93  &	    \\
{\rm A~1878} &  14h & 13m  & 00.40s &	29d & 13m  & 37.50s &             &  -19.13  &  -18.71  &	    \\
{\rm A~1878} &  14h & 12m  & 57.01s &	29d & 13m  & 43.90s &             &  -19.27  &  -18.62  &	    \\
{\rm A~1878} &  14h & 12m  & 57.70s &	29d & 13m  & 48.90s &             &  -19.78  &  -19.18  &   S0      \\
{\rm A~1878} &  14h & 13m  & 03.99s &	29d & 13m  & 53.50s &             &  -19.36  &  -19.74  &	    \\
{\rm A~1878} &  14h & 13m  & 02.65s &	29d & 14m  & 01.20s &             &  -18.08  &  -18.25  &	    \\
{\rm A~2111} &  15h & 39m  & 35.52s &	34d & 26m  & 56.20s &             &  -20.46  &  -19.56  &    S     \\
{\rm A~2111} &  15h & 39m  & 37.64s &	34d & 27m  & 03.80s &  0.2295     &  -21.26  &  -20.22  &   S0      \\
{\rm A~2111} &  15h & 39m  & 31.84s &	34d & 27m  & 05.10s &             &  -19.01  &  -18.09  &	    \\
{\rm A~2111} &  15h & 39m  & 38.48s &	34d & 24m  & 32.40s &             &  -19.37  &  -18.96  &	    \\
{\rm A~2111} &  15h & 39m  & 40.16s &	34d & 24m  & 18.10s &             &  -19.38  &  -18.33  &	    \\
{\rm A~2111} &  15h & 39m  & 39.34s &	34d & 24m  & 44.50s &             &  -20.26  &  -19.31  &    E      \\
{\rm A~2111} &  15h & 39m  & 38.45s &	34d & 24m  & 51.40s &             &  -20.02  &  -19.08  &	    \\
{\rm A~2111} &  15h & 39m  & 40.16s &	34d & 24m  & 55.70s &             &  -20.49  &  -19.27  &   S0      \\
{\rm A~2111} &  15h & 39m  & 42.76s &	34d & 24m  & 56.60s &             &  -18.32  &  -17.38  &	    \\
{\rm A~2111} &  15h & 39m  & 37.84s &	34d & 24m  & 57.00s &             &  -18.55  &  -17.54  &	    \\
{\rm A~2111} &  15h & 39m  & 39.81s &	34d & 25m  & 00.50s &             &  -18.06  &  -17.79  &	    \\
{\rm A~2111} &  15h & 39m  & 37.21s &	34d & 25m  & 08.50s &             &  -19.67  &  -18.70  &    S     \\
{\rm A~2111} &  15h & 39m  & 40.49s &	34d & 25m  & 27.30s &  0.2282     &  -22.67  &  -21.51  &    E      \\
{\rm A~2111} &  15h & 39m  & 39.75s &	34d & 25m  & 23.10s &             &  -19.57  &  -18.85  &	    \\
{\rm A~2111} &  15h & 39m  & 39.20s &	34d & 25m  & 11.50s &             &  -21.13  &  -20.38  &    E     \\
{\rm A~2111} &  15h & 39m  & 39.39s &	34d & 25m  & 13.40s &  0.2211     &  -21.34  &  -20.61  &    E      \\
{\rm A~2111} &  15h & 39m  & 36.23s &	34d & 25m  & 12.10s &             &  -20.34  &  -19.21  &   S0      \\
{\rm A~2111} &  15h & 39m  & 34.90s &	34d & 25m  & 14.50s &             &  -18.97  &  -18.02  &	    \\
{\rm A~2111} &  15h & 39m  & 40.27s &	34d & 25m  & 34.80s &             &  -20.07  &  -21.06  &	    \\
{\rm A~2111} &  15h & 39m  & 38.15s &	34d & 25m  & 18.10s &             &  -20.21  &  -19.01  &   S0      \\
{\rm A~2111} &  15h & 39m  & 37.53s &	34d & 25m  & 18.70s &             &  -19.62  &  -18.72  &    S      \\
{\rm A~2111} &  15h & 39m  & 36.64s &	34d & 25m  & 29.00s &             &  -18.96  &  -18.34  &	    \\
{\rm A~2111} &  15h & 39m  & 33.61s &	34d & 25m  & 34.00s &             &  -18.48  &  -17.78  &	   \\
{\rm A~2111} &  15h & 39m  & 36.79s &	34d & 25m  & 39.10s &  0.2312     &  -20.90  &  -19.65  &   S0      \\
{\rm A~2111} &  15h & 39m  & 39.69s &	34d & 25m  & 21.20s &             &  -20.40  &  -19.70  &	    \\
{\rm A~2111} &  15h & 39m  & 31.27s &	34d & 25m  & 40.00s &             &  -20.24  &  -19.65  &	    \\
{\rm A~2111} &  15h & 39m  & 38.68s &	34d & 25m  & 38.90s &             &  -18.16  &  -17.04  &	    \\
{\rm A~2111} &  15h & 39m  & 37.29s &	34d & 25m  & 45.90s &             &  -18.49  &  -17.33  &	    \\
{\rm A~2111} &  15h & 39m  & 36.42s &	34d & 25m  & 50.10s &             &  -20.46  &  -19.50  &    E      \\
{\rm A~2111} &  15h & 39m  & 41.20s &	34d & 25m  & 50.90s &             &  -20.53  &  -19.36  &   S0      \\
{\rm A~2111} &  15h & 39m  & 40.18s &	34d & 25m  & 50.80s &             &  -18.60  &  -17.90  &	    \\
{\rm A~2111} &  15h & 39m  & 33.99s &	34d & 25m  & 51.30s &             &  -19.30  &  -18.23  &	    \\
{\rm A~2111} &  15h & 39m  & 37.44s &	34d & 25m  & 54.80s &             &  -20.12  &  -18.98  &    E      \\
{\rm A~2111} &  15h & 39m  & 39.52s &	34d & 25m  & 56.90s &             &  -19.23  &  -18.23  &	    \\
{\rm A~2111} &  15h & 39m  & 39.91s &	34d & 25m  & 57.20s &             &  -18.53  &  -17.69  &	    \\
{\rm A~2111} &  15h & 39m  & 41.69s &	34d & 26m  & 01.70s &             &  -18.03  &  -16.89  &	    \\
{\rm A~2111} &  15h & 39m  & 31.72s &	34d & 26m  & 07.20s &             &  -20.49  &  -19.37  &   S0      \\
{\rm A~2111} &  15h & 39m  & 36.84s &	34d & 26m  & 07.20s &             &  -20.44  &  -19.68  &    I      \\
{\rm A~2111} &  15h & 39m  & 38.07s &	34d & 26m  & 09.50s &             &  -18.64  &  -19.00  &	    \\
{\rm A~2111} &  15h & 39m  & 34.11s &	34d & 26m  & 19.20s &             &  -20.80  &  -20.58  &    S      \\
{\rm A~2111} &  15h & 39m  & 34.26s &	34d & 26m  & 12.50s &  0.2289     &  -21.97  &  -21.11  &   S0     \\
{\rm A~2111} &  15h & 39m  & 38.18s &	34d & 26m  & 06.90s &             &  -19.72  &  -19.53  &	    \\
{\rm A~2111} &  15h & 39m  & 32.26s &	34d & 26m  & 12.80s &             &  -19.25  &  -18.22  &	    \\
{\rm A~2111} &  15h & 39m  & 38.58s &	34d & 26m  & 28.20s &             &  -20.23  &  -19.24  &    S      \\
{\rm A~2111} &  15h & 39m  & 39.03s &	34d & 26m  & 38.10s &             &  -19.29  &  -19.48  &	    \\
{\rm A~2111} &  15h & 39m  & 38.70s &	34d & 26m  & 38.80s &  0.2246     &  -20.85  &  -20.29  &    S     \\
{\rm A~2111} &  15h & 39m  & 37.81s &	34d & 26m  & 35.90s &             &  -18.12  &  -17.73  &	    \\
{\rm A~2111} &  15h & 39m  & 31.99s &	34d & 26m  & 36.10s &             &  -18.39  &  -18.00  &	    \\
{\rm A~2111} &  15h & 39m  & 35.47s &	34d & 26m  & 43.70s &             &  -20.70  &  -19.87  &   S0      \\
{\rm A~2111} &  15h & 39m  & 41.19s &	34d & 26m  & 41.30s &             &  -20.27  &  -20.24  &    I      \\
{\rm A~2111} &  15h & 39m  & 40.90s &	34d & 26m  & 45.40s &             &  -19.28  &  -19.45  &	    \\
{\rm A~2111} &  15h & 39m  & 37.59s &	34d & 26m  & 44.20s &             &  -18.92  &  -18.91  &	    \\
{\rm A~2111} &  15h & 39m  & 33.13s &	34d & 26m  & 45.60s &             &  -19.25  &  -18.91  &	    \\
{\rm A~2111} &  15h & 39m  & 37.16s &	34d & 26m  & 45.70s &             &  -18.26  &  -17.93  &	    \\
{\rm A~2111} &  15h & 39m  & 38.38s &	34d & 26m  & 50.50s &             &  -18.01  &  -17.13  &	    \\
{\rm A~2111} &  15h & 39m  & 32.78s &	34d & 24m  & 22.40s &             &  -19.21  &  -18.31  &	    \\
{\rm A~2111} &  15h & 39m  & 41.34s &	34d & 24m  & 34.30s &  0.2294     &  -20.97  &  -20.81  &    S      \\
{\rm A~2111} &  15h & 39m  & 41.81s &	34d & 24m  & 42.70s &  0.2292     &  -22.61  &  -22.18  &    E     \\
{\rm A~2111} &  15h & 39m  & 42.27s &	34d & 24m  & 40.40s &             &  -19.08  &  -20.81  &	   \\
{\rm A~2111} &  15h & 39m  & 41.26s &	34d & 24m  & 43.60s &             &  -20.43  &  -22.04  &    S      \\
{\rm A~2111} &  15h & 39m  & 47.09s &	34d & 27m  & 37.90s &  0.2368     &  -21.25  &  -20.57  &   S0      \\
{\rm A~2111} &  15h & 39m  & 42.81s &	34d & 27m  & 44.60s &             &  -19.68  &  -19.55  &    I     \\
{\rm A~2111} &  15h & 39m  & 52.99s &	34d & 27m  & 48.60s &  0.2297     &  -20.98  &  -19.94  &   S0      \\
{\rm A~2111} &  15h & 39m  & 54.29s &	34d & 25m  & 06.60s &             &  -18.05  &  -17.24  &	   \\
{\rm A~2111} &  15h & 39m  & 51.92s &	34d & 25m  & 18.80s &             &  -18.85  &  -18.83  &	    \\
{\rm A~2111} &  15h & 39m  & 44.40s &	34d & 25m  & 22.70s &             &  -19.46  &  -19.41  &	    \\
{\rm A~2111} &  15h & 39m  & 44.15s &	34d & 25m  & 21.30s &             &  -18.78  &  -20.66  &	    \\
{\rm A~2111} &  15h & 39m  & 54.03s &	34d & 25m  & 24.60s &             &  -18.87  &  -18.31  &	    \\
{\rm A~2111} &  15h & 39m  & 53.10s &	34d & 25m  & 26.50s &             &  -18.75  &  -17.73  &	    \\
{\rm A~2111} &  15h & 39m  & 47.96s &	34d & 25m  & 32.10s &             &  -20.49  &  -19.52  &    E      \\
{\rm A~2111} &  15h & 39m  & 52.98s &	34d & 25m  & 41.10s &             &  -19.31  &  -18.24  &	    \\
{\rm A~2111} &  15h & 39m  & 43.94s &	34d & 25m  & 46.70s &             &  -19.45  &  -19.01  &	    \\
{\rm A~2111} &  15h & 39m  & 42.69s &	34d & 25m  & 52.10s &             &  -18.13  &  -16.99  &	    \\
{\rm A~2111} &  15h & 39m  & 42.04s &	34d & 26m  & 04.00s &             &  -19.44  &  -18.39  &	    \\
{\rm A~2111} &  15h & 39m  & 42.02s &	34d & 25m  & 59.60s &             &  -18.96  &  -20.78  &	    \\
{\rm A~2111} &  15h & 39m  & 44.85s &	34d & 25m  & 58.50s &             &  -19.19  &  -19.08  &	    \\
{\rm A~2111} &  15h & 39m  & 47.82s &	34d & 26m  & 00.00s &             &  -19.08  &  -18.05  &	    \\
{\rm A~2111} &  15h & 39m  & 53.40s &	34d & 25m  & 59.60s &             &  -18.94  &  -18.56  &	    \\
{\rm A~2111} &  15h & 39m  & 52.50s &	34d & 26m  & 02.20s &             &  -20.40  &  -19.62  &    S      \\
{\rm A~2111} &  15h & 39m  & 47.49s &	34d & 26m  & 11.10s &             &  -18.84  &  -18.00  &	    \\
{\rm A~2111} &  15h & 39m  & 42.59s &	34d & 26m  & 14.00s &             &  -19.82  &  -18.96  &	    \\
{\rm A~2111} &  15h & 39m  & 43.06s &	34d & 26m  & 23.00s &             &  -18.42  &  -17.36  &	    \\
{\rm A~2111} &  15h & 39m  & 42.02s &	34d & 26m  & 30.30s &  0.2258     &  -22.09  &  -20.96  &    E      \\
{\rm A~2111} &  15h & 39m  & 43.25s &	34d & 26m  & 32.90s &             &  -19.02  &  -18.04  &	   \\
{\rm A~2111} &  15h & 39m  & 48.31s &	34d & 26m  & 36.70s &             &  -18.79  &  -18.43  &	    \\
{\rm A~2111} &  15h & 39m  & 49.35s &	34d & 26m  & 41.50s &  0.2299     &  -21.54  &  -21.03  &    S      \\
{\rm A~2111} &  15h & 39m  & 50.11s &	34d & 26m  & 44.40s &             &  -19.48  &  -19.02  &	   \\
{\rm A~2111} &  15h & 39m  & 52.85s &	34d & 26m  & 46.80s &             &  -20.45  &  -19.95  &    E      \\
{\rm A~2111} &  15h & 39m  & 42.09s &	34d & 26m  & 49.20s &             &  -19.55  &  -18.52  &	    \\
{\rm A~2111} &  15h & 39m  & 45.75s &	34d & 26m  & 57.40s &  0.2292     &  -21.07  &  -20.05  &    E      \\
{\rm A~2111} &  15h & 39m  & 42.98s &	34d & 27m  & 00.30s &             &  -18.40  &  -18.28  &	   \\
{\rm A~2111} &  15h & 39m  & 42.30s &	34d & 27m  & 02.60s &             &  -18.20  &  -17.92  &	    \\
{\rm A~2111} &  15h & 39m  & 52.55s &	34d & 27m  & 07.50s &             &  -19.06  &  -20.58  &	    \\
{\rm A~2111} &  15h & 39m  & 52.04s &	34d & 27m  & 07.60s &             &  -19.01  &  -21.29  &	    \\
{\rm A~2111} &  15h & 39m  & 52.15s &	34d & 27m  & 12.20s &             &  -21.13  &  -21.23  &    S      \\
{\rm A~2111} &  15h & 39m  & 42.28s &	34d & 27m  & 17.10s &             &  -20.12  &  -19.17  &	    \\
{\rm A~2111} &  15h & 39m  & 51.51s &	34d & 27m  & 31.30s &             &  -19.54  &  -18.61  &	    \\
{\rm A~2111} &  15h & 39m  & 48.27s &	34d & 27m  & 34.80s &             &  -18.70  &  -18.45  &	    \\
{\rm A~2111} &  15h & 39m  & 47.70s &	34d & 25m  & 16.40s &             &  -19.85  &  -18.96  &	    \\
{\rm A~2111} &  15h & 39m  & 47.89s &	34d & 27m  & 39.90s &             &  -20.28  &  -19.32  &    E      \\
{\rm A~2111} &  15h & 39m  & 47.34s &	34d & 25m  & 10.20s &  0.2309     &  -21.07  &  -20.81  &    E      \\
{\rm A~2111} &  15h & 39m  & 47.26s &	34d & 25m  & 15.90s &             &  -20.43  &  -20.38  &    S     \\
{\rm A~1952} &  14h & 41m  & 07.84s &	28d & 38m  & 29.40s &             &  -22.05  &  -21.10  &    E      \\
{\rm A~1952} &  14h & 40m  & 59.08s &	28d & 38m  & 35.40s &             &  -20.11  &  -19.24  &    S      \\
{\rm A~1952} &  14h & 41m  & 01.82s &	28d & 35m  & 57.10s &             &  -20.13  &  -19.75  &    S      \\
{\rm A~1952} &  14h & 40m  & 59.60s &	28d & 36m  & 07.40s &             &  -19.18  &  -18.44  &	    \\
{\rm A~1952} &  14h & 41m  & 02.64s &	28d & 36m  & 14.50s &             &  -18.79  &  -18.40  &	    \\
{\rm A~1952} &  14h & 41m  & 01.57s &	28d & 36m  & 31.50s &             &  -18.30  &  -18.20  &	    \\
{\rm A~1952} &  14h & 40m  & 59.42s &	28d & 36m  & 42.00s &             &  -19.05  &  -18.36  &	    \\
{\rm A~1952} &  14h & 41m  & 04.07s &	28d & 36m  & 47.50s &             &  -19.94  &  -19.04  &    E      \\
{\rm A~1952} &  14h & 41m  & 04.47s &	28d & 36m  & 49.70s &             &  -18.71  &  -20.59  &	    \\
{\rm A~1952} &  14h & 41m  & 01.82s &	28d & 37m  & 09.60s &             &  -18.17  &  -20.22  &	    \\
{\rm A~1952} &  14h & 41m  & 01.92s &	28d & 37m  & 14.50s &             &  -20.76  &  -20.80  &    E      \\
{\rm A~1952} &  14h & 41m  & 02.66s &	28d & 37m  & 10.00s &             &  -22.11  &  -21.94  &   S0      \\
{\rm A~1952} &  14h & 41m  & 03.13s &	28d & 37m  & 10.10s &             &  -21.41  &  -20.84  &    E      \\
{\rm A~1952} &  14h & 41m  & 02.67s &	28d & 37m  & 02.40s &             &  -20.63  &  -19.99  &	    \\
{\rm A~1952} &  14h & 40m  & 58.41s &	28d & 36m  & 52.50s &             &  -19.89  &  -19.03  &   S0      \\
{\rm A~1952} &  14h & 41m  & 01.19s &	28d & 37m  & 00.50s &             &  -21.20  &  -20.33  &    E      \\
{\rm A~1952} &  14h & 40m  & 59.94s &	28d & 37m  & 22.10s &             &  -18.59  &  -17.95  &	    \\
{\rm A~1952} &  14h & 40m  & 59.55s &	28d & 37m  & 34.20s &             &  -18.29  &  -17.81  &	    \\
{\rm A~1952} &  14h & 41m  & 01.81s &	28d & 37m  & 34.70s &             &  -19.48  &  -18.91  &	    \\
{\rm A~1952} &  14h & 41m  & 01.58s &	28d & 37m  & 48.30s &             &  -18.21  &  -20.37  &	    \\
{\rm A~1952} &  14h & 41m  & 01.32s &	28d & 37m  & 43.20s &             &  -21.57  &  -21.82  &    E      \\
{\rm A~1952} &  14h & 41m  & 01.53s &	28d & 37m  & 44.30s &             &  -19.21  &  -18.96  &	    \\
{\rm A~1952} &  14h & 40m  & 59.15s &	28d & 37m  & 47.80s &             &  -18.93  &  -20.67  &	    \\
{\rm A~1952} &  14h & 40m  & 59.50s &	28d & 37m  & 48.80s &             &  -19.90  &  -19.01  &	    \\
{\rm A~1952} &  14h & 40m  & 58.98s &	28d & 37m  & 51.40s &             &  -18.41  &  -18.51  &	    \\
{\rm A~1952} &  14h & 41m  & 03.17s &	28d & 37m  & 52.50s &             &  -18.10  &  -17.71  &	    \\
{\rm A~1952} &  14h & 40m  & 59.94s &	28d & 38m  & 00.10s &             &  -20.38  &  -19.53  &    E     \\
{\rm A~1952} &  14h & 41m  & 08.82s &	28d & 37m  & 59.00s &             &  -19.72  &  -19.04  &    E     \\
{\rm A~1952} &  14h & 41m  & 05.82s &	28d & 38m  & 02.20s &             &  -18.68  &  -17.82  &	   \\
{\rm A~1952} &  14h & 41m  & 00.90s &	28d & 38m  & 04.70s &             &  -19.54  &  -18.59  &	   \\
{\rm A~1952} &  14h & 41m  & 04.06s &	28d & 38m  & 08.40s &             &  -19.13  &  -19.23  &	   \\
{\rm A~1952} &  14h & 41m  & 07.98s &	28d & 38m  & 09.40s &             &  -19.08  &  -18.20  &	   \\
{\rm A~1952} &  14h & 41m  & 03.17s &	28d & 38m  & 21.10s &             &  -18.04  &  -18.15  &	   \\
{\rm A~1952} &  14h & 41m  & 05.43s &	28d & 38m  & 21.80s &             &  -19.11  &  -18.99  &	   \\
{\rm A~1952} &  14h & 41m  & 02.63s &	28d & 35m  & 50.80s &             &  -18.52  &  -18.54  &	    \\
{\rm A~1952} &  14h & 40m  & 59.43s &	28d & 38m  & 27.20s &             &  -19.17  &  -19.37  &	    \\
{\rm A~1952} &  14h & 41m  & 01.91s &	28d & 35m  & 54.80s &             &  -19.06  &  -20.83  &	    \\
{\rm A~1952} &  14h & 40m  & 59.20s &	28d & 38m  & 24.20s &             &  -19.32  &  -18.60  &	    \\
{\rm A~1952} &  14h & 41m  & 13.59s &	28d & 37m  & 29.60s &             &  -22.13  &  -21.43  &   S0      \\
{\rm A~1952} &  14h & 41m  & 05.84s &	28d & 37m  & 41.60s &             &  -20.22  &  -19.27  &	    \\
{\rm A~1952} &  14h & 41m  & 14.94s &	28d & 37m  & 42.60s &             &  -21.80  &  -20.98  &   S0      \\
{\rm A~1952} &  14h & 41m  & 05.68s &	28d & 37m  & 46.70s &             &  -18.22  &  -18.33  &	    \\
{\rm A~1952} &  14h & 41m  & 08.53s &	28d & 37m  & 49.00s &             &  -19.26  &  -19.18  &	    \\
{\rm A~1952} &  14h & 41m  & 03.16s &	28d & 37m  & 52.20s &             &  -18.08  &  -17.86  &	    \\
{\rm A~1952} &  14h & 41m  & 15.18s &	28d & 35m  & 29.10s &             &  -18.16  &  -20.08  &	    \\
{\rm A~1952} &  14h & 41m  & 15.04s &	28d & 35m  & 21.20s &             &  -19.32  &  -21.52  &	    \\
{\rm A~1952} &  14h & 41m  & 15.18s &	28d & 35m  & 24.30s &             &  -19.61  &  -21.59  &	    \\
{\rm A~1952} &  14h & 41m  & 03.94s &	28d & 35m  & 21.30s &             &  -20.35  &  -19.38  &    E      \\
{\rm A~1952} &  14h & 41m  & 13.55s &	28d & 35m  & 21.80s &             &  -19.77  &  -19.72  &	    \\
{\rm A~1952} &  14h & 41m  & 05.92s &	28d & 35m  & 29.90s &             &  -20.77  &  -19.82  &   S0      \\
{\rm A~1952} &  14h & 41m  & 08.36s &	28d & 35m  & 28.50s &             &  -18.46  &  -18.64  &	    \\
{\rm A~1952} &  14h & 41m  & 08.59s &	28d & 35m  & 30.80s &             &  -19.01  &  -19.42  &	    \\
{\rm A~1952} &  14h & 41m  & 08.51s &	28d & 35m  & 32.50s &             &  -20.41  &  -20.52  &    S      \\
{\rm A~1952} &  14h & 41m  & 04.76s &	28d & 35m  & 32.40s &             &  -19.40  &  -19.16  &	    \\
{\rm A~1952} &  14h & 41m  & 07.83s &	28d & 35m  & 32.30s &             &  -18.39  &  -20.48  &	    \\
{\rm A~1952} &  14h & 41m  & 07.59s &	28d & 35m  & 35.00s &             &  -21.58  &  -21.23  &    S      \\
{\rm A~1952} &  14h & 41m  & 11.01s &	28d & 35m  & 33.00s &             &  -19.87  &  -18.93  &	    \\
{\rm A~1952} &  14h & 41m  & 10.10s &	28d & 35m  & 33.70s &             &  -19.03  &  -18.15  &	    \\
{\rm A~1952} &  14h & 41m  & 05.91s &	28d & 35m  & 38.50s &             &  -19.36  &  -18.50  &	    \\
{\rm A~1952} &  14h & 41m  & 08.19s &	28d & 35m  & 44.50s &             &  -21.61  &  -20.93  &   S0      \\
{\rm A~1952} &  14h & 41m  & 03.56s &	28d & 35m  & 44.20s &             &  -19.69  &  -19.45  &	    \\
{\rm A~1952} &  14h & 41m  & 13.72s &	28d & 35m  & 54.10s &             &  -18.37  &  -18.45  &	    \\
{\rm A~1952} &  14h & 41m  & 13.55s &	28d & 35m  & 52.00s &             &  -18.78  &  -21.38  &	    \\
{\rm A~1952} &  14h & 41m  & 03.27s &	28d & 35m  & 56.30s &             &  -19.34  &  -18.51  &	    \\
{\rm A~1952} &  14h & 41m  & 08.00s &	28d & 36m  & 03.20s &             &  -18.96  &  -18.97  &	    \\
{\rm A~1952} &  14h & 41m  & 09.73s &	28d & 36m  & 02.80s &             &  -18.20  &  -17.44  &	    \\
{\rm A~1952} &  14h & 41m  & 05.67s &	28d & 36m  & 05.30s &             &  -18.64  &  -17.72  &	    \\
{\rm A~1952} &  14h & 41m  & 12.15s &	28d & 36m  & 07.00s &             &  -19.45  &  -18.73  &	    \\
{\rm A~1952} &  14h & 41m  & 10.72s &	28d & 36m  & 07.50s &             &  -18.10  &  -18.28  &	    \\
{\rm A~1952} &  14h & 41m  & 14.47s &	28d & 36m  & 26.20s &             &  -19.86  &  -19.18  &	    \\
{\rm A~1952} &  14h & 41m  & 05.45s &	28d & 36m  & 26.40s &             &  -18.40  &  -17.58  &	    \\
{\rm A~1952} &  14h & 41m  & 04.06s &	28d & 36m  & 26.50s &             &  -18.06  &  -20.09  &	    \\
{\rm A~1952} &  14h & 41m  & 04.22s &	28d & 36m  & 27.80s &             &  -18.35  &  -19.37  &	    \\
{\rm A~1952} &  14h & 41m  & 12.94s &	28d & 36m  & 27.30s &             &  -19.29  &  -18.61  &	    \\
{\rm A~1952} &  14h & 41m  & 06.81s &	28d & 36m  & 31.50s &             &  -20.01  &  -21.90  &    E      \\
{\rm A~1952} &  14h & 41m  & 07.10s &	28d & 36m  & 37.30s &             &  -20.67  &  -20.27  &    E      \\
{\rm A~1952} &  14h & 41m  & 07.03s &	28d & 36m  & 39.20s &             &  -22.10  &  -22.55  &   S0      \\
{\rm A~1952} &  14h & 41m  & 03.36s &	28d & 36m  & 37.10s &             &  -20.43  &  -19.40  &    E      \\
{\rm A~1952} &  14h & 41m  & 14.04s &	28d & 36m  & 40.80s &             &  -18.80  &  -18.92  &	    \\
{\rm A~1952} &  14h & 41m  & 03.11s &	28d & 36m  & 46.60s &             &  -20.74  &  -19.93  &   S0      \\
{\rm A~1952} &  14h & 41m  & 04.07s &	28d & 36m  & 52.70s &             &  -19.32  &  -20.90  &	    \\
{\rm A~1952} &  14h & 41m  & 03.57s &	28d & 37m  & 00.30s &             &  -22.61  &  -24.23  &    E      \\
{\rm A~1952} &  14h & 41m  & 03.14s &	28d & 36m  & 57.00s &             &  -19.55  &  -21.16  &	    \\
{\rm A~1952} &  14h & 41m  & 10.75s &	28d & 36m  & 47.30s &             &  -19.16  &  -21.67  &	    \\
{\rm A~1952} &  14h & 41m  & 06.34s &	28d & 37m  & 01.30s &             &  -18.75  &  -18.04  &	    \\
{\rm A~1952} &  14h & 41m  & 06.48s &	28d & 37m  & 06.90s &             &  -20.14  &  -19.23  &	    \\
{\rm A~1952} &  14h & 41m  & 08.25s &	28d & 37m  & 13.80s &             &  -21.85  &  -21.21  &    S      \\
{\rm A~1952} &  14h & 41m  & 12.33s &	28d & 37m  & 11.00s &             &  -19.03  &  -19.11  &	   \\
{\rm A~1952} &  14h & 41m  & 06.26s &	28d & 37m  & 12.20s &             &  -18.15  &  -17.28  &	   \\
{\rm A~1952} &  14h & 41m  & 09.40s &	28d & 37m  & 13.00s &             &  -18.55  &  -18.59  &	   \\
{\rm A~1952} &  14h & 41m  & 09.72s &	28d & 37m  & 17.80s &             &  -19.85  &  -20.07  &    S     \\
{\rm A~1952} &  14h & 41m  & 05.02s &	28d & 37m  & 34.90s &             &  -18.05  &  -20.08  &	   \\
{\rm A~1952} &  14h & 41m  & 06.27s &	28d & 37m  & 27.50s &             &  -20.01  &  -20.84  &   S0     \\
{\rm A~1952} &  14h & 41m  & 05.49s &	28d & 37m  & 33.90s &             &  -19.02  &  -21.47  &	   \\
{\rm A~1952} &  14h & 41m  & 05.32s &	28d & 37m  & 35.90s &             &  -19.18  &  -20.34  &	   \\
{\rm A~1952} &  14h & 41m  & 04.78s &	28d & 37m  & 31.60s &             &  -19.79  &  -21.64  &	   \\
{\rm A~1952} &  14h & 41m  & 04.76s &	28d & 37m  & 35.50s &             &  -19.83  &  -19.94  &	   \\
{\rm A~1952} &  14h & 41m  & 04.77s &	28d & 35m  & 05.50s &             &  -18.30  &  -18.55  &	    \\
{\rm A~1952} &  14h & 41m  & 06.32s &	28d & 37m  & 18.30s &             &  -18.30  &  -17.78  &	    \\
{\rm A~1952} &  14h & 41m  & 14.57s &	28d & 37m  & 18.70s &             &  -19.65  &  -19.37  &	    \\
{\rm A~1952} &  14h & 41m  & 11.82s &	28d & 37m  & 19.30s &             &  -18.39  &  -18.59  &	    \\
{\rm A~1952} &  14h & 41m  & 07.53s &	28d & 37m  & 23.60s &             &  -19.14  &  -18.20  &	    \\
{\rm A~1952} &  14h & 41m  & 04.29s &	28d & 37m  & 23.00s &             &  -18.54  &  -18.27  &	    \\
{\rm A~1952} &  14h & 41m  & 05.23s &	28d & 35m  & 06.80s &             &  -19.10  &  -18.75  &	    \\
{\rm A~1952} &  14h & 41m  & 03.35s &	28d & 37m  & 29.50s &             &  -20.17  &  -19.26  &	    \\
{\rm A~1952} &  14h & 41m  & 06.39s &	28d & 37m  & 33.60s &             &  -18.12  &  -17.60  &	    \\
{\rm A~1952} &  14h & 41m  & 12.23s &	28d & 35m  & 06.70s &             &  -19.05  &  -18.27  &	    \\
{\rm A~1952} &  14h & 41m  & 08.44s &	28d & 35m  & 09.00s &             &  -18.39  &  -17.78  &	    \\
{\rm A~1952} &  14h & 41m  & 05.36s &	28d & 37m  & 40.50s &             &  -18.42  &  -17.71  &	    \\
{\rm A~1952} &  14h & 41m  & 10.78s &	28d & 35m  & 12.30s &             &  -18.24  &  -18.05  &	    \\
{\rm A~1952} &  14h & 41m  & 13.01s &	28d & 35m  & 15.30s &             &  -19.08  &  -19.23  &	    \\
{\rm A~1952} &  14h & 41m  & 10.45s &	28d & 35m  & 16.60s &             &  -19.36  &  -19.26  &	    \\
{\rm A~1952} &  14h & 41m  & 14.88s &	28d & 35m  & 29.70s &             &  -18.45  &  -19.98  &	    \\

\hline
\label{tab:dcat}
\end{longtable}

\end{appendix}

\end{document}